\documentclass[12pt]{iopart}

\usepackage{iopams}  
\usepackage{perpage}

\newcommand{\E}{e}
\newcommand{\bea}{\begin{eqnarray}}
\newcommand{\eea}{\end{eqnarray}}
\newcommand{\beq}{\begin{eqnarray}}
\newcommand{\eeq}{\end{eqnarray}}
\newcommand{\de}{\partial}
\newcommand{\ba}{\begin{eqnarray}}
\newcommand{\ea}{\end{eqnarray}}

\newcommand{\MPd}{M_{P}^2}

\newcommand{\MPmd}{M_{P}^{-2}}
\newcommand{\be}{\begin{equation}}
\newcommand{\ee}{\end{equation}}

\def\L5{\tilde{\Lambda}}

\begin{document}

\title[An introduction to the Vainshtein mechanism]{An introduction to the Vainshtein mechanism}

\author{Eugeny Babichev$^1$ and C\'edric Deffayet$^2$}

\address{$^1$ Laboratoire de Physique Th\'eorique d'Orsay,
B\^atiment 210, Universit\'e Paris-Sud 11,
F-91405 Orsay Cedex, France.}

\address{$^2$ APC (UMR 7164 - APC, Univ Paris Diderot, CNRS/IN2P3, CEA/lrfu, Obs de Paris, 
Sorbonne Paris CitÈ, France), 10 rue Alice Domon et L\'eonie Duquet, 75205 Paris Cedex 13, France.}

\eads{\mailto{eugeny.babichev@th.u-psud.fr}, \mailto{deffayet@iap.fr}} 


\begin{abstract}
We introduce the Vainshtein mechanism which plays a crucial role in massive gravities, as well as in related theories such as Galileons and their extensions. This mechanism, also known as k-mouflage, allows to hide via non linear effects -- typically for source distances smaller than a so-called Vainshtein radius which depends on the source and on the theory considered -- some degrees of freedom whose effects are then only left important at large distances, e.g. for cosmology. It is introduced here in non linear Fierz-Pauli theories (massive gravities), including the dRGT theories, in their decoupling limits, as well as in other models such as DGP model or generalized Galileons. This presentation is self-contained and before discussing the Vainshtein mechanism we introduce some useful results and concepts concerning massive gravity, such as the vDVZ discontinuity, the decoupling limits or the Boulware-Deser ghost.

\end{abstract}

\maketitle

\section{Introduction}
The idea to give a mass to the graviton is not new and has been investigated by many authors since the first years of General Relativity (for reviews see this volume and \cite{Rubakov:2008nh,Hinterbichler:2011tt}). 
It recently regained popularity after the invention of the Dvali-Gabadadze-Porrati (DGP) model \cite{Dvali:2000hr} and the discovery of its interesting cosmology \cite{Deffayet:2000uy,Deffayet:2001pu}. Indeed, on the one hand, this model was the first where a large distance modification of gravity was shown to lead to cosmic acceleration even in the absence of a cosmological constant; on the other  hand, the DGP model shares many properties which can be expected from a theory of massive gravity. The connection between the DGP model and massive gravity can of course be related to the fact that, in  the DGP model, gravity is just mediated by a continuum of massive Kaluza-Klein gravitons due to the flat and higher dimensional character of the bulk space-time. It is also amusing to note that there  appears to be a historical connection between  the very introduction of the cosmological constant (and hence the standard explanation of the later discovery of the accelerating expansion \cite{SCP,HZT}) and the idea to give a mass to the graviton (Einstein hoping that a cosmological constant would induce what is nowadays called a Yukawa decay of the gravitational potential \cite{SCHU}). 

The simplest theory for a non self-interacting  massive graviton is known as the Fierz-Pauli theory \cite{Pauli:1939xp,Fierz:1939ix}. It suffers
 from a pathology known as the vDVZ discontinuity \cite{Veltman} as will be introduced below. This is enough to rule out such a theory from basic solar system tests of gravity. However, soon after the discovery of the vDVZ discontinuity, a way out was suggested by Vainshtein \cite{Arkady}, relying on a non linear extension of the Fierz-Pauli theory. This proposal of Vainshtein was however lacking a proper proof, as was underlined in particular in Ref.~\cite{Boulware:1973my} which appeared soon after Vainshtein's paper in 1972. The later reference also discovered another pathology of the non linear Fierz-Pauli theories, the so-called Boulware-Deser ghost, that we will also introduce below. The situation remained unchanged until the advent of the DGP model where the Vainshtein mechanism was re-introduced (the DGP model featuring also the vDVZ discontinuity), and where new arguments were given in favour of the validity of this mechanism \cite{Deffayet:2001uk}. Following this, a number of works tried to prove or disprove the Vainshtein mechanism in DGP or other simpler models related to massive gravity  (see references in the following). 
 The purpose of this work is to introduce with some detail this mechanism and discuss to which extend it can be considered established in the light of these recent developments, stressing that this mechanism plays a crucial role for phenomenological applications of massive gravity or of its close friends (such as Galileons, covariant Galileons and their generalizations \cite{Nicolis:2008in,Deffayet:2009wt,Horn,Fairlie}). 
  
We first, in sections \ref{Genres} and \ref{NLFP}, introduce massive gravity theories: Fierz-Pauli theory, its simplest non linear generalizations and their generic properties. The results presented in these sections \ref{Genres} and \ref{NLFP} are mostly known for more than 40 years (with the exception of the decoupling limit and associated strong coupling introduced in section \ref{SCDL}) and serve as a background for the following. In the way we present them, however, we will use some more recent works which allowed to discuss them in a simple and clear way.
We then, in section \ref{NEWMASS} present more recent developments: the DGP model and the de Rham-Gabadadze-Tolley (dRGT) theory  of massive gravity \cite{deRham:2010kj,deRham:2010ik,deRham:2011rn}. The latter theory was built specifically not to suffer from  the presence of the Boulware-Deser ghost, a pathology that was once thought to be unavoidable in simple non linear extensions of Fierz-Pauli models (dRGT theory is  presented in section \ref{dRGT}). In section \ref{Vain}, we introduce the Vainshtein mechanism and the recent progresses that have been made about its understanding in the various theories introduced before.

\section{Massive gravities, generalities}

 \label{Genres}

\subsection{Fierz-Pauli theory}
\label{PFsec}
Fierz-Pauli theory\footnote{In this work, we will keep the name "Fierz-Pauli theory" to describe the quadratic theory introduced in this subsection. This theory has linear field equations. By contrast, we will call "Non-Linear Fierz-Pauli theory" -- henceforth NLFP -- a non linear completion of the later theory of the type introduced in the following section \ref{NLFP}.} is the only consistent non self-interacting Lorentz invariant theory for a massive spin 2 \cite{Pauli:1939xp,Fierz:1939ix}. It can be defined by the following action defined on flat space-time with canonical metric $g_{\mu \nu} = \eta_{\mu \nu}$
\ba \label{PF}
S_{PF} & =& \MPd \int d^4x   \left[ -\frac{1}{4}\left(\partial_\mu h_{\nu\rho} \right)^2 + \frac{1}{4}\left(\partial_\mu h\right)^2 \nonumber
- \frac{1}{2} \left(\partial_\mu h\right)\left( \partial^\nu h^\mu_\nu \right)
\right.\\
&&\left.+ \frac{1}{2} \left(\partial_\mu h_{\nu \rho}\right) \left(\partial^\nu h^{\mu \rho}\right)
-\frac{1}{4}m^2\left( h_{\mu \nu}h^{\mu \nu}-h^2\right) +\MPmd T_{\mu \nu}h^{\mu \nu}\right]
\ea 
where  $h_{\mu \nu}$ is some rank-2 covariant tensor, $m$ and $M_P$ are mass parameters, the indices of $h_{\mu \nu}$ are moved up and down with the metric $\eta_{\mu \nu}$ and $h=h_{\mu \nu}\eta^{\mu \nu}$, while $h_{\mu \nu}$ is coupled to some energy momentum tensor $T_{\mu \nu}$. The graviton $h_{\mu \nu}$ has a mass term in the Fierz-Pauli form, that one reads from the above equation as given by 
\ba \label{PFmass}
S_{PF,m} & =& -  \frac{1}{4} \MPd m^2 \int d^4x \left( h_{\mu \nu}h^{\mu \nu}-h^2\right).
\ea 
Any other combination of $h_{\mu \nu}h^{\mu \nu}$ and $h^2$ would lead to instabilities \cite{Pauli:1939xp,Fierz:1939ix}. Note also that this mass term explicitly breaks gauge invariance.
The first four terms of action (\ref{PF}) are simply obtained by expanding at quadratic order into $h_{\mu \nu}$ the Einstein-Hilbert action around flat space-time.
Varying action (\ref{PF}) with respect to $h_{\mu \nu}$, we get the following equation of motion 
\ba \label{FIELDgen}
{\cal E}_{\mu \nu} = -\frac{1}{2}m^2 \left(h_{\mu \nu} - h \eta_{\mu \nu}\right) + \MPmd T_{\mu \nu}, 
\ea
where ${\cal E}_{\mu \nu}$ is the linearization around $\eta_{\mu \nu}$ of the Einstein tensor ${G}_{\mu \nu}$ and is given by 
\ba
{\cal E}_{\mu \nu} &=&
\mathcal{E}^{\alpha\beta}_{\mu\nu}h_{\alpha\beta}  \nonumber \\
&=& -\frac{1}{2} \partial_\mu \partial_\nu h - \frac{1}{2} \Box h_{\mu \nu} + \frac{1}{2} \partial_\rho \partial_\mu h^\rho_\nu 
+ \frac{1}{2}\partial_\rho \partial_\nu h^\rho_\mu  - \frac{1}{2} \eta_{\mu \nu}(\partial^\rho \partial^\sigma h_{\rho \sigma} - \Box h) \nonumber
\ea
It is easily checked that, as a consequence of Bianchi identities, ${\cal E}_{\mu \nu}$ is divergenceless. Hence, if one further assumes that $T_{\mu \nu}$ is conserved with respect to the background derivative $\partial$, i.e. that one has $\partial^\mu T_{\mu \nu} =0$,
we get (taking the divergence of the field equations ({\ref{FIELDgen})) for non vanishing graviton mass $m$
\ba \label{transversegen}
\partial^\rho h_{\rho \mu} &=&  \partial_\mu h.
\ea
This represents four first-order equations which eliminate four degrees
of freedom out of the ten a priori independent components of $h_{\mu \nu}$.
Taking one more derivative of the above equation, we have  
\ba \label{hmnh}
 \partial^\nu \partial^\mu h_{\mu \nu} - \Box h = 0.
\ea
The left hand side of the last equation is nothing but the linearized Ricci scalar. Eq. (\ref{hmnh}) can be fruitfully used in the trace (with respect to $\eta_{\mu \nu}$) of the field equations (\ref{FIELDgen}), to get 
\ba \label{hTflat}
h = -\frac{2}{3}\frac{T}{m^2 M_P^2}.
\ea
Hence the trace $h$ of $h_{\mu \nu}$ is determined algebraically and does not propagate. We will use this equation below, but just notice at this point that, if one considers the theory in vacuum, equations  (\ref{transversegen}) and (\ref{hTflat}) imply that a massive Fierz-Pauli graviton in vacuum is transverse and traceless, and hence contains 5 degrees of freedom (the same conclusion can be reached by a rigorous hamiltonian counting, as we will recall later).

\subsection{The vDVZ discontinuity}
Plugging (\ref{transversegen})  and (\ref{hTflat}) into Eq. (\ref{FIELDgen}) we get easily that 
\begin{equation}
-\frac{1}{2}\left(\Box -  m^2\right) h_{\mu \nu} 
= \frac{1}{M_P^{2}} \left(T_{\mu \nu} - \frac{1}{3} T \eta_{\mu \nu}\right) 
+  \frac{1}{3} \frac{\partial_\mu \partial_\nu T}{m^2 M_P^2}.
\label{hmT}
\end{equation}
Using then a Fourier decomposition of $h_{\mu \nu}$ (and using similar notations for $T_{\mu \nu}$)
\begin{equation}
h_{\mu \nu} (x^\rho) 
=  \frac{1}{(2 \pi)^4} \int d^4 k\; e^{i k^\mu x_\mu} \bar{h}_{\mu \nu} \left(k^\rho \right),
\end{equation}
we get from (\ref{hmT}) the expression of the propagator (in Fourier space) $\bar{D}^{(m\neq0)}_{\mu \nu \rho \sigma}$, such that 
\begin{equation}
\bar{h}_{\mu \nu} = \bar{D}^{(m\neq0)}_{\mu \nu \rho \sigma} \frac{\bar{T}^{\rho \sigma}}{M_P^2} ,
\end{equation}
given by 
\begin{equation}
\label{propagm}
\bar{D}^{(m\neq0)}_{\mu \nu \rho \sigma} = \frac{1}{k^2 +  m^2}  \left(\eta_{\rho \mu} \eta_{\sigma \nu} + \eta_{\rho \nu} \eta_{\sigma \mu}-\frac{2}{3} \eta_{\rho \sigma} \eta_{\mu \nu}  - \frac{2}{3} \eta_{\rho \sigma} \frac{k_\mu k_\nu}{m^2} \right),
\end{equation}
where $k^2 = k^\mu k_\mu$.
This should be compared to the propagator of a massless graviton $\bar{D}^{(0)}_{\mu \nu \rho \sigma}$ which reads in Fourier space 
\begin{equation}
\label{propagm0}
\bar{D}^{(m=0)}_{\mu \nu \rho \sigma} \sim \frac{1}{k^2}  \left(\eta_{\rho \mu} \eta_{\sigma \nu} + \eta_{\rho \nu} \eta_{\sigma \mu}- \eta_{\rho \sigma} \eta_{\mu \nu} \right),
\end{equation}
where the symbol $\sim$ means that gauge dependent and momentum dependent terms are omitted. 
Notice the crucial difference between expressions (\ref{propagm}) and (\ref{propagm0}) lying in the third term in the parentheses on the right hand side of these expressions. This difference in the coefficient in front of $\eta_{\rho \sigma} \eta_{\mu \nu}$ is independent of the mass of the graviton and is at the root of the so-called vDVZ discontinuity (vDVZ standing for "van Dam-Veltman-Zakharov") which states that, roughly speaking, however small the graviton mass, Fierz-Pauli theory leads to different physical predictions (such as light bending) from those of linearized General Relativity \cite{Veltman}.  There are various ways to see this. Let us here (following e.g. \cite{Porrati:2000cp}) consider the formal tree level amplitude ${\cal A}$ between two conserved current $T_{\mu \nu}$ and $S_{\mu \nu}$, defined as  
\begin{equation}
{\cal A} = M_P^{2}\int d^4 x \; S^{\mu \nu}(x) h_{\mu \nu}\left(T\right)(x),
\end{equation}
where $h_{\mu \nu}\left(T\right)$ is the tree level graviton field generated by the conserved source $T_{\rho \sigma}$, and given by
\begin{equation}
h_{\mu \nu}(T)(x)= M_P^{-2}\int d^4 x'\;{D}_{\mu \nu \rho \sigma}\left(x-x'\right)T^{\rho \sigma}\left(x'\right),
\end{equation}
(${D}_{\mu \nu \rho \sigma}$ being the massless or massive propagator).
The amplitude ${\cal A}$ is easily obtained in Fourier space, using the expressions (\ref{propagm}) and (\ref{propagm0}) for the propagators. 
One  gets respectively in the massless and massive cases, and in the large $k$ limit ($k \gg m$) 
\ba
{\cal A}^{(m = 0)} &=& \int d^4 k \; \frac{2}{k^2}\left(\bar{S}^{\mu \nu}\bar{T}_{\mu \nu}- \frac{1}{2} \bar{S}\bar{T}\right) \\
{\cal A}^{(m\neq0)} &=& \int d^4 k \; \frac{2}{k^2}\left(\bar{S}^{\mu \nu}\bar{T}_{\mu \nu}- \frac{1}{3} \bar{S}\bar{T}\right).
\ea
 Considering then non relativistic currents --  such that 
$ \bar{T}^\mu_{\nu} \propto {\rm diag}(\bar{M_1},0,0,0) $ and 
$ \bar{S}^\mu_{\nu} \propto {\rm diag}(\bar{M_2},0,0,0)$ --  separated by a distance  small with respect to the graviton Compton wavelength (which diverges as $m$ goes to zero), the amplitude due to the exchange of a massive graviton is given approximately by 
\begin{equation}
\label{AMPLIT}
{\cal A}^{(m \neq 0)} = \frac{4}{3} {\cal A}^{(m=0)} = \frac{4}{3} \int d^4 k \frac{\bar{M}_1 \bar{M}_2}{k^2} ,
\end{equation}
so that the massive amplitude stays different from the massless one, however small the graviton mass. 
For the same non relativistic sources, this translates into a similar
discrepancy in the potentials, the potential of the massive theory being
larger by a factor 4/3, and it reflects an extra attraction in the massive
theory with respect to the massless theory.
This extra attraction can be attributed to the exchange of the helicity zero polarization of the massive graviton (which, as we recalled above, has 3 more polarizations than the massless one). It can be eliminated by redefining the Newton constant of the massive theory with respect to the massless one,
assuming, e.g., that one measures the Newton constant by some Cavendish experiment (indeed, if one does not do such a rescaling, the Newton constant of the massive theory would be given by  $4/3 \times M_P^{-2}/ 16 \pi = M_P^{-2}/ 12 \pi$). However, with such a rescaling, the discontinuity will then reappear in other observables, like the light bending. The latter will then be 25\% smaller in the massive case than in the massless one \cite{Veltman}, which is much too large to be compatible with current measurements of the light bending by the sun \cite{Beringer:1900zz,Will:2005va}. 

As such, the vDVZ discontinuity is enough to rule out from standard solar system tests of gravity any theory where it appears\footnote{Note however that the discontinuity does not appear if the background is maximally symmetric with a non vanishing curvature \cite{Porrati:2000cp,Higushi,Kogan1} and that this result can be extended to more general situations \cite{Deffayet:2004ru,Koyama:2007ih,Fasiello:2012rw}.  However, depending on the value of the mass of the graviton, the theory can be non unitary when background curvature does not vanish \cite{Higushi}.}. It must be cured if one wants to make such a theory viable. The Vainshtein mechanism (first introduced in Ref. \cite{Arkady}) allows in principle to get rid of the vDVZ discontinuity, as will be explained in the following. It relies on non linearities in the field equations which are absent (by definition) in the Fierz-Pauli theory. It is also clear that a gravity theory able to approach General Relativity in high curvature regimes must be non linear, independently of the Vainshtein mechanism, and there are then  several good reasons  to consider non linear completion of the Fierz-Pauli theory. 

Historically, the non linear completion that played the most important role for the recent developments about massive gravity, is certainly the Dvali-Gabadadze-Porrati model (henceforth DGP model) \cite{Dvali:2000hr}. This model will be briefly discussed below in section \ref{DGP}, and, as we said,  it is this model and its cosmological consequences, that, in the years 2000, lead to a very strong renewal of interest about massive gravity, related theories (such as, e.g., Galileons \cite{Nicolis:2008in}) and the Vainshtein mechanism (as discussed in \cite{Deffayet:2001uk}). However, the DGP model is far from being the simplest non linear extension of Fierz-Pauli model one can consider, since in particular it contains a non countable infinity of massive gravitons. We introduce in the next section a  family of much simpler non linear completions, that we will call here and henceforth "Non Linear Fierz-Pauli" theories or NLFP. The Vainshtein mechanism was originaly introduced in a theory close to one of this family\footnote{It was pointed out in \cite{Damour:2002gp} that the theory used by Vainshtein does not fall strictly speaking into this class, but it does not matter for the discussions of this paper.} and the recently discussed massive gravity of de Rham, Gabadadze and Tolley \cite{deRham:2010kj,deRham:2010ik,deRham:2011rn} also belongs to this family. Note that the NLFP theories were first discussed in relation with strong interactions \cite{Isham:gm}.

\section{Non Linear Fierz Pauli gravity}

\label{NLFP}

\subsection{Action and equations of motion}
An obvious way to generalize in a non linear way the Fierz-Pauli theory considered in the previous section, as well as to stay close to General Relativity (henceforth GR), is to consider a theory with a dynamical metric $g_{\mu \nu}$ and {\it the same} kinetic term as the one of GR, hence given by the usual Einstein-Hilbert action reading 
\begin{equation}
S_{EH} = 
\frac{\MPd}{2}\left(\int d^4 x \sqrt{-g} \left(R-2\Lambda \right)\right)\label{kinEH},
\end{equation}
with a possibly non vanishing cosmological constant $\Lambda$, and to add to this term a mass term. This mass term should be such that when expanded at quadratic order around some suitable background for $g_{\mu \nu}$, namely that given by a flat metric $\eta_{\mu \nu}$,  it reduces to the Fierz-Pauli form (\ref{PFmass}) where $h_{\mu \nu}$ is now viewed as the dynamical field representing the fluctuation of the metric $g_{\mu \nu}$ around its background expression. Ideally, the mass term for $g_{\mu \nu}$ should only depend on $g_{\mu \nu}$ itself in a non derivative way. However, the only such non trivial term corresponds to a Lagrangian density proportional to the volume element $\sqrt{-g}$, hence to a cosmological constant. Moreover, such a term, if expanded around some arbitrary background metric contains a tadpole (in particular when the background metric is the Minkowski metric), and does not have the Fierz-Pauli form at quadratic order. In fact, it is clear that it does not give any mass to the graviton since this term does not break general covariance and hence leads to a theory with two propagating degrees of freedom. 

Hence, the sought for mass term requires the introduction of some extra field besides the metric $g_{\mu \nu}$. One possibility we shall consider here is to introduce an extra metric $f_{\mu \nu}$ that will be taken to be flat and non dynamical\footnote{Note that one can also consider the extra metric $f_{\mu \nu}$ to cover a non flat space-time such as de Sitter or Anti de Sitter}. As will be explained below, such a theory can {\it equivalently} be seen as that of a dynamical metric $g_{\mu \nu}$ and four scalar fields. The two metrics $f_{\mu \nu}$ and $g_{\mu \nu}$ will be taken to be non derivatively coupled via an interaction term  $S_{int}[f,g]$. This term will be chosen such that (i) the theory is general covariant under diffeomorphisms (common to the two metrics), (ii) it has flat space-time as a solution of the field equations for $g_{\mu \nu}$, and (iii) that when one expands $g_{\mu \nu}$ to second order around the canonical Minkowski metric $\eta_{\mu\nu}$
as $g_{\mu\nu} = \eta_{\mu\nu} + h_{\mu\nu}$ and let $f_{\mu \nu}$ to have the canonical Minkowski form $\eta_{\mu \nu}$, the potential at quadratic order for $h_{\mu \nu}$ takes the Fierz-Pauli form (\ref{PFmass}).  There is much freedom in the choice of such
an interaction term. For example, the following two possibilities have be considered respectively 
by Boulware and Deser  in Ref. \cite{Boulware:1973my} and  by Arkani-Hamed {\it et al.} in Ref. \cite{Arkani-Hamed:2002sp}
\ba
S_{int}^{(2)}&\equiv&-\frac{1}{8} m^2 M_{P}^{2}\int d^{4}x \; \sqrt{-f}\; H_{\mu \nu} H_{\sigma \tau}\left(f^{\mu\sigma}f^{\nu\tau}-f^{\mu\nu}f^{\sigma\tau}\right) \label{S2} \\
S_{int}^{(3)}&\equiv&-\frac{1}{8} m^2 M_{P}^{2}\int d^{4}x \; \sqrt{-g}\;H_{\mu \nu} H_{\sigma \tau}\left(g^{\mu\sigma}g^{\nu\tau}-g^{\mu\nu}g^{\sigma\tau}\right), \label{S3} 
\ea
where $f^{\mu \nu}$ and $g^{\mu \nu}$ denotes respectively the inverse of the metric $f_{\mu \nu}$ and $g_{\mu \nu}$,  and  $H_{\mu \nu}$ is defined by $ 
H_{\mu \nu} = g_{\mu \nu} - f_{\mu \nu} $. More generally, one can consider theories where the interaction term $S_{int}$ is not of the above forms (\ref{S2}-\ref{S3})\footnote{We keep the numbering of the interaction terms (\ref{S2}-\ref{S3}) of Ref. \cite{Damour:2002ws}.}, 
but is chosen such that it obeys properties (i), (ii) and (iii) above.
We will later introduce the interaction terms specific to the dRGT theories \cite{deRham:2010kj,deRham:2010ik,deRham:2011rn} which also differ from the ones above.
Note that since we have two metrics at hand, there is some  ambiguity on how to move  indices up and down. Here, when necessary, indices will be moved with the dynamical metric $g$ except for indices of the other metric $f_{\mu \nu}$ itself. I.e. $f^{\mu \nu}$ is defined as the inverse of the metric $f_{\mu \nu}$ and hence it is {\it not} given by $g^{\mu \sigma} g^{\nu \rho} f_{\sigma \rho}$ (this means in particular that $H^{\mu \nu}$ is not given by $g^{\mu \nu}-f^{\mu \nu}$).
Following the notations of Damour {\it et al.} \cite{Damour:2002ws}, one notes that the interaction terms considered so far all have the form
\be \label{INT}
S_{int}^{(a)} = -\frac{1}{8} m^2 M_{P}^{2} \int d^{4}x {\cal V}^{(a)} (g,f) \equiv  -\frac{1}{8} m^2 M_{P}^{2} \int d^{4}x \sqrt{-g}V^{(a)} ({\bf g^{-1} f})
\ee
with ${\cal V}^{(a)}(g,f)\equiv \sqrt{-g}\;V^{(a)}({\bf g^{-1} f})$ a suitable "potential" density associated with the scalar function $V^{(a)}$. In fact one can show that this form follows necessarily from the assumption that the metric 
$f$ and $g$ interact in a non derivative way in the interaction term $S_{int}[f,g]$, and that the theory is invariant under diffeomorphism \cite{Damour:2002ws}. This later invariance is represented as usual as the transformations acting on the metrics of the following form
\begin{equation}\label{Ginvg}
\eqalign{
g_{\mu \nu}(x) &= \partial_\mu x'^{\sigma}(x) \partial_\nu x'^{\tau}(x) g'_{\sigma \tau}\left(x'(x)\right)\;, \\
f_{\mu \nu}(x) &= \partial_\mu x'^{\sigma}(x) \partial_\nu x'^{\tau}(x) f'_{\sigma \tau}\left(x'(x)\right)\;, }
\end{equation}
and under which the quantity $V^{(a)}$ transforms as a scalar. 
 Introducing matter, there is also much freedom as far as choosing the metric to which matter couples. Indeed, e.g.,  an 
infinite family of inequivalent metric can be built from the two metric at hand $f_{\mu \nu}$ and $g_{\mu \nu}$, and one could decide to couple matter minimally to one arbitrary metric in this family. Here, we will consider the simplest case where matter is assumed to be minimally coupled to the metric $g$, hence the total action of the theory we will consider here is given by 
\begin{equation}
\label{action} S=\int d^4 x\sqrt{-g}\ \left(\frac{M_P^2}{2} R +L[g]\right)  + S_{int}[f,g],
\end{equation}
where in the above action, $L[g]$ denotes a generic matter Lagrangian with a minimal coupling to $g_{\mu \nu}$ (and not to the metric $f_{\mu \nu}$), and we have included a possibly non vanishing cosmological constant in the interaction term $S_{int}[f,g]$.

The field equations, derived from action (\ref{action}), read
\be \label{EQMot}
M_P^2 G_{\mu \nu} =\left(T_{\mu \nu}+ T^g_{\mu \nu}\right),
\ee
where $G_{\mu\nu}$ denotes
the Einstein tensor computed with the metric $g$,
$T_{\mu \nu}$ is the energy momentum tensor of matter fields, and
$T^g_{\mu \nu}$ is the effective energy momentum tensor coming from the variation with respect to the metric $g$ of the interaction term $S_{int}$. It depends non derivatively on both metrics $f$ and $g$ and is defined as usual as
\be \label{DEFTMN}
T^{g}_{\mu \nu}(x) = - \frac{2}{\sqrt{-g}} \frac{\delta}{\delta g^{\mu \nu}(x)} S_{int}[f,g].
\ee
It is then easy to check with these expressions that the equations of motion (\ref{EQMot}) reduce indeed to the Fierz-Pauli equations (\ref{FIELDgen}) at linearized level. A simple, but non trivial, consequence of equations (\ref{EQMot}) is obtained by taking a g-covariant derivative $\nabla$ of both sides of the equations; one gets, using the Bianchi identities and the conservation of the matter energy momentum tensor, the constraint
\be \label{BIAN}
\nabla^\mu T_{\mu \nu}^g =0
\ee
which the effective energy momentum tensor should obey. 

We will call a theory having an action of the form (\ref{action}) discussed above (where we recall that the interaction term obeys conditions (i), (ii) and (iii) above) a Non Linear Fierz-Pauli (NLFP) theory. We will now introduce a pathology, first discussed by Boulware and Deser \cite{Boulware:1973my}, and once thought to be present in any NLFP theory.

\subsection{The Boulware-Deser ghost}
\label{BDghost}
It is here convenient to contrast a Hamiltonian analysis of the Fierz-Pauli theory with one of a generic NLFP theory, the starting point being a $3+1$ decomposition of the graviton or of the metric.

Since the kinetic term of the linear Fierz-Pauli theory is the same as that of linearized General Relativity, one can use first well known results concerning GR. In particular, one sees that neither $h_{00}$ nor $h_{0i}$ (with $i=1,2,3$ spatial indices) are dynamical degrees of freedom since their canonical momentum vanish identically in massive gravity as well as for a massless graviton and $h_{00}$ and $h_{0i}$ are Lagrange multipliers in the kinetic term obtained by expanding the Einstein-Hilbert action. In the mass term, however, $h_{00}$ and $h_{0i}$ play quite different r\^oles. Indeed, the mass term (\ref{PFmass}) reads 
\beq 
-\frac{1}{4} M_P^2 m^2 \label{Massterm}
\int d^4 x \left\{ h_{ij} h_{ij} - 2 h_{0i}h_{0i} - h_{ii}h_{jj}+ 2 h_{ii}h_{00}\right\}, 
\eeq
and it hence appears that $h_{00}$ is a Lagrange multiplier for the entire action since it also enters linearly in the mass term. As a consequence, the field equation for $h_{00}$ generates a constraint which reads  
\beq \label{CONSHOO}
\nabla^2 h_{ii}-h_{ij, ij} \propto m^2 h_{ii} , 
\eeq 
and allows to eliminate one extra degree of freedom\footnote{To state things more rigorously, on can show that this constraint is second class and generates an extra constraint of the same class.  These constraints  together kill two Hamiltonian degrees of freedom, i.e. one Lagrangian degree of freedom.}. In contrast, the field equations for the $h_{0i}$, which appear quadratically in the mass term, do not eliminate any degree of freedom (in contrast to what happens in the massless case), they determine the $h_{0i}$ in terms of the other dynamical components. Together, this leaves a total of 5 propagating degrees of freedom, in agreement with the discussion given above in section \ref{PFsec}.

Let us now see how those results are modified in the case of a NLFP theory. In the massless case, that is to say for General Relativity formulated {\it \`a la}  ADM  \cite{Arnowitt:1962hi}, Lagrange multipliers associated with diffeomorphism invariance are the  "lapse" $N$ and "shifts" $N^{i}$, respectively defined as  $ N\equiv 1/ \sqrt{-g^{00}} $ and  $ N_i 
\equiv g_{0i}$  in terms of the components of the metric
$g_{\mu \nu}$. They generate (first class) constraints which eliminate 4 out of the 6 possible dynamical degrees of freedom, leaving the well-known 2 polarizations of a massless graviton. The addition of a NLFP "mass term" such as  (\ref{S2})-(\ref{S3}) modifies however notably the nature of $N$ and $N^{i}$ (as it breaks invariance under diffeomorphisms). For example, the action with the mass term  (\ref{S2}) reads in the first order formalism (after a convenient renormalisation of $m^2$)
\ba &&
M_P^2\int d^4 x \left\{ \left(\pi^{ij}\dot{g}_{ij}
- N R^0 - N_i R^i\right)\right. \nonumber\\
&& \left.\;\;\;\;\;\;\; - m^2 \left(h_{ij}h_{ij} - 2 N_i N_i -
h_{ii} h_{jj} + 2 h_{ii} \left(1-N^2 + N_k g^{kl} \label{EHADM}
N_l\right)\right) \right\}, 
\ea 
where the $\pi^{ij}$ are conjugate momenta to the $g_{ij}$ and $R^0$ and $R^i$ are respectively the Hamiltonian and momentum constraints of GR (generated by the lapse and shifts) and cubic terms simply follow from the standard definitions of $N$ and $N_i$ recalled above.
As Boulware and Deser first pointed out \cite{Boulware:1973my}  neither  
$N_i$ nor $N$ are Lagrange multipliers of the non linear theory. Hence, the number of propagating degrees of freedom is generically 6 and not 5. Boulware and Deser also argued that the reduced Hamiltonian for the 6 physical degrees of freedom  is in general unbounded from below, and this can indeed easily be checked explicitly in some cases such as that of action (\ref{EHADM}) (see e.g. \cite{Deffayet:2005ys}). Given the unboundedness-from-below nature of the Hamiltonian the extra-mode is usually called the "Boulware-Deser ghost".  Note also that (as discussed e.g. in \cite{Deffayet:2005ys}) one can also understand the presence of the Boulware-Deser ghost in a covariant way using the observation that the constraint  (\ref{hTflat}) is lost in generic  NLFP theories and is replaced (at cubic order), schematically, by a quadratic equation of the form 
 \beq
  \Box h^2 +m^2 h \propto M_P^{-2} T
 \eeq
 (while a constraint such as (\ref{transversegen}) still holds in the form of (\ref{BIAN})). Note that this discussion is modified for dRGT theories which were precisely built to eliminate the BD ghost  but note also that this does not exhaust  all the possible pathologies (or their cures) of non linear massive gravity (see e.g. \cite{Gruzinov:2011sq,Burrage:2011cr,Burrage:2012ja,Deser:2012qx,Deser:2013eua,Babichev:2013una}). 

\subsection{Strong coupling and decoupling limit}
\label{SCDL}
It was first noticed in \cite{Deffayet:2001uk}, in the context of the DGP model, that the Vainshtein mechanism can only work at the price of a strong coupling in the considered theories. A subsequent work \cite{Arkani-Hamed:2002sp} introduced a powerful method to extract from a generic NLFP theory a simple theory which captures many of the crucial properties concerning this strong coupling and its relation with the Vainshtein mechanism. The purpose of this subsection is to introduce this method and some of its outcomes. 

A good starting point is to notice that the gauge invariance (\ref{Ginvg}) can be used to write the background flat metric $f$ in various coordinate systems. Starting from  a given gauge, with coordinate $X^A$,
 and the $f$ metric in the form of $f_{AB}(X)$, it might be desirable to change the gauge, but keep the change of coordinate explicit in the $f$ metric. Namely, the action considered takes the form of (\ref{action}), but with $f_{\mu \nu}(x)$ now given by  the expression
\ba \label{STUCA}
f_{\mu \nu}(x) &=& \partial_\mu X^A(x) \partial_\nu X^B(x) f_{AB}\left(X(x)\right),
\ea
while $g$ is kept as $g_{\mu \nu}(x)$. The quantities $X^A$, which then appear explictly in the action of the theory,
can be considered as a set of four new dynamical scalar fields, which are analogous to the Stuckelberg field used to restore gauge invariance in the Proca Lagrangian \cite{Arkani-Hamed:2002sp,Dubovsky:2004sg}.

With this in mind, the initial gauge, where $g$ and $f$ assume the form $g_{AB}$ and $f_{AB}$ is usually called a "unitary gauge", i.e. one where the Stuckelberg fields $X^A$ are gauged away. Note that the metric $f_{\mu \nu}$ in a non unitary gauge  can also be thought as the pullback, via the "link field" $X^A(x)$, on the space-time manifold $m_4$, with coordinates $x^\mu$, of the metric $f_{AB}$ living in an other abstract manifold ${\cal M}_4$ with coordinates $X^A$ \cite{Arkani-Hamed:2002sp}.
Usually, the unitary gauge is chosen such that in this gauge and when the extra metric $f$ is that of a flat space-time, $f_{AB}$ takes the canonical Minkowski form $\eta_{AB} \equiv {\rm diag}(-1,1,1,1)$. In the non unitary gauge, the action  (\ref{action}) is one for a theory with $g_{\mu \nu}$ and $X^A$ as dynamical fields. Obviously, the equations of motion for $g^{\mu \nu}$ lead to the same equations as in (\ref{EQMot}) where $f_{\mu \nu}$ is given the form (\ref{STUCA}).
On the other hand, it is not difficult to show that the field equations for the $X^A$ are equivalent to the Bianchi identities (\ref{BIAN}) provided that the mapping $X^A(x)$ is invertible (for an explicit proof see e.g. \cite{Babichev:2009us}). 

The authors of Ref.~\cite{Arkani-Hamed:2002sp} have further developed the above mentioned analogy between $X^A$ and Stuckelberg fields doing a "Goldstone boson" expansion of the action 
 (\ref{action}) around a unitary gauge. Considering some background solution for $g_{\mu \nu}$ (defined as $g_{\mu \nu}^0$) and $X^A(x)$ defined (the metric $f_{AB}$ being kept fixed) as 
\be
X_0^A(x) \equiv \delta^A_\mu x^\mu,\nonumber
\ee
Ref. \cite{Arkani-Hamed:2002sp} introduces the "pion" fields $\pi^A$ as 
\be \label{DECPIPI}
X^A(x) = X^A_0(x) + \pi^A(x),
\ee
and further does a "scalar-vector" decomposition of the $\pi^A$ in the form\footnote{This form is in fact not explicitely the one given in  \cite{Arkani-Hamed:2002sp} but it seems to be what is  done there implicitely. See \cite{Babichev:2009us} for a discussion about the associated subtleties which do in fact matter when one discusses the terms beyond quadratic order.}
\be \label{DECPIbis}
\pi^A(x) = \delta^A_\mu\left(A^\mu(x)+ \eta^{\mu \nu} \partial_\nu \phi\right).
\ee
The above equation introduces new fields $A^\mu$ and $\phi$ but also associated gauge symmetries, and hence, in line with the original idea of St\"{u}ckelberg, this does not change the number of 
(propagating) degrees of freedom, but rather reshuffles them in a different way.
If one inserts this decomposition  into action (\ref{action}), and expands around flat space-time writing 
$g_{\mu \nu} = \eta_{\mu \nu} + h_{\mu \nu}$, we obtain an action for the dynamical fields $h_{\mu \nu}(x)$, $A^{\mu}(x)$ and $\phi(x)$. Since $A^\mu(x)$ and $\phi(x)$ only enter in the metric $f_{\mu \nu}$ (via expression (\ref{STUCA})) the only term in action  (\ref{action}) which depends on $A^\mu(x)$ and $\phi(x)$ is the interaction term $S_{int}[f,g]$, and one has 
 (where no term has been neglected in the expression below)
\begin{equation}
\label{EXPANDH}
\eqalign{
H_{\mu \nu} &= h_{\mu \nu} - \partial_\mu A_\nu - \partial_\nu A_\mu - 2 \partial_\mu \partial_\nu \phi  - \partial_\mu A_\sigma \partial_\nu A^\sigma \\
&- \partial_\mu \partial_\sigma \phi \; \partial_\nu \partial^\sigma \phi  - \partial_\nu A^\sigma \partial_\mu \partial_\sigma \phi - \partial_\mu A^\sigma \partial_\nu \partial_\sigma \phi.
}
\end{equation}
Following again \cite{Arkani-Hamed:2002sp}, we can obtain canonically normalized fields $\tilde{\phi}$, $\tilde{A}$ and $\hat{h}_{\mu\nu}$ by defining 
 \be
 \label{PHIRESCA}
 \eqalign{
 \hat{h}_{\mu \nu} =  M_P {h}_{\mu \nu}, \;
\tilde{A}^\mu = M_P m A^\mu, \;
\tilde{\phi} = M_P m^2 \phi. 
}
\ee
Inserting (\ref{EXPANDH}) with (\ref{PHIRESCA}) into $S_{int}[f,g]$, keeping the lowest order in $h_{\mu \nu}(x)$, $A^\mu(x)$ and $\phi(x)$, 
and using the redefinition 
\beq \hat h_{\mu \nu} = \tilde{h}_{\mu \nu} - \eta_{\mu \nu} \tilde\phi, \label{defhhatht} \eeq
 one obtains the following quadratic action 
\begin{equation} \label{qexpbis}
\label{qexp}
\eqalign{
S =\frac{1}{8}\int d^{4}x &
\Big\{ 2 \tilde{h}^{\mu\nu} \partial_{\mu}\partial_{\nu}\tilde{h} - 2 \tilde{h}^{\mu \nu} \partial_\nu \partial_\sigma \tilde{h}^\sigma_\mu + \tilde{h}^{\mu \nu} 
 \Box \tilde{h}_{\mu\nu} - \tilde{h} \Box \tilde{h}\\
& + m^2 \left(\tilde{h}^{2}-\tilde{h}_{\mu\nu}\tilde{h}^{\mu\nu}\right) - \tilde{F}_{\mu\nu}\tilde{F}^{\mu\nu} -4m\left(\tilde{h}\partial \tilde{A}-\tilde{h}_{\mu\nu}\partial^{\mu}\tilde{A}^{\nu}\right)\\
&+6\tilde{\phi}(\Box+2m^{2})\tilde{\phi}-m^2 \tilde{h}\phi+2m\tilde{\phi}\partial\tilde{A} \Big\}
+\frac{1}{2}\frac{T_{\mu\nu}}{M_P}\tilde{h}^{\mu\nu} -\frac{1}{2} \frac{T}{M_P} \tilde{\phi}\nonumber \\
}
\end{equation}
where $\tilde{h} \equiv \tilde{h}_{\mu \nu} \eta^{\mu \nu}$, $\partial\tilde{A} \equiv \partial_\mu\tilde{A}^\mu$, $T_{\mu\nu}$ is the matter stress-energy tensor, $T=T^{\mu \nu} \eta_{\mu \nu}$, and indices are moved up and down with the metric $\eta_{\mu \nu}$.
The peculiarity of the above action is that while $\tilde{A}^\mu$ acquired directly a standard kinetic term, $\tilde\phi$ did only 
via a mixing with $\hat{h}_{\mu \nu}$ (that was demixed through definition (\ref{defhhatht})) \cite{Arkani-Hamed:2002sp}, this being entirely due to the structure of the Pauli-Fierz mass term (\ref{PFmass}). The remaining cross terms between $\tilde\phi$, $\tilde A$ and $\tilde{h}_{\mu\nu}$ can be cancelled by adding an appropriate gauge fixing to the action (see \cite{Nibbelink:2006sz}).
 
Expanding the action in $\tilde{\phi}$, $\tilde{A}$ and $\tilde{h}_{\mu\nu}$ to next orders, one sees  that
$\tilde{\phi}$ has in general cubic self interactions suppressed by the energy scale
\be \label{DEFLAMBDA}
\Lambda_5 = \left(m^4 M_P\right)^{1/5}\;.
\ee
When these interactions are present\footnote[1]{An appropriate choice of the interaction term $S_{inf}[f,g]$ can remove cubic (and some others) self interactions of $\tilde{\phi}$ \cite{Arkani-Hamed:2002sp}.}, they are the strongest interactions among the fields $\tilde{\phi}$, $\tilde{A}$ and $\tilde{h}_{\mu\nu}$ in the limit where $m \ll M_P$. One can take a {\it decoupling limit} (henceforth DL) defined as 
\be\label{DEFDEC}
M_{P}\rightarrow \infty,\; m  \rightarrow  0,\;  \Lambda_5 \sim {\rm constant},\;  T_{\mu \nu}/M_P  \sim {\rm constant},
\ee 
in order to isolate this interaction. In this limit, the action one is left with for $\tilde{\phi}$ is of the form 
\be\label{action_phi}
S=\frac{1}{2}\int d^{4}x\;\left\{\frac{3}{2}\tilde{\phi}\Box\tilde{\phi}+\frac{1}{\Lambda_5^{5}}\left[\alpha \;(\Box\tilde{\phi})^{3}+\beta\;(\Box\tilde{\phi}\;\tilde{\phi}_{,\mu\nu}\;\tilde{\phi}^{,\mu\nu})\right]-\frac{1}{M_{P}}T\tilde{\phi}\right\}\,,
\ee
where $\alpha$ and $\beta$ are numerical coefficients that depend on the interaction term $S_{inf}[f,g]$\footnote{Note that in general, the cubic term for $\tilde{\phi}$ is given by some linear combination of the three terms 
$(\Box\tilde{\phi})^{3}$, $\Box\tilde{\phi}\;\tilde{\phi}_{,\mu\nu}\;\tilde{\phi}^{,\mu\nu}$  and $\tilde{\phi}_{,\mu\nu}\;\tilde{\phi}^{,\mu\sigma}\tilde{\phi}_{,\sigma}^{,\nu}$, but an integration by parts can always be used to reduce the number of independent terms to two, as shown in Eq. (\ref{action_phi}).}.  
For example, the potential (\ref{S2}) leads to $\alpha=-\beta=-1/2$, while the  potential (\ref{S3}) leads to the opposite case $\alpha=-\beta=1/2$. In contrast, in the DL, the other fields $\tilde{h}_{\mu \nu}$ and $\tilde{A}^\mu$ are just free. 

The equation of motion deriving from action (\ref{action_phi}) is 
\be
3\Box\tilde{\phi}+\frac{1}{\Lambda^{5}}\left[3\alpha \;\Box\left(\Box\tilde{\phi}\right)^{2}+\beta\;\Box\left(\tilde{\phi}_{,\mu\nu}\;\tilde{\phi}^{,\mu\nu}\right)+2\beta\;\partial_{\mu}\partial_{\nu}\left(\Box\tilde{\phi}\;\tilde{\phi}^{,\mu\nu}\right)\right]=\frac{1}{M_{P}}T\; .
\label{tildephi}
\ee
As will be discussed in section~\ref{KMOUFLAGE} the Vainshtein mechanism can be easily read off from this equation. Moreover, the fact that this equation is of fourth order signals that action (\ref{action_phi}) propagates in fact two scalar modes, one being ghost like. This ghost can be interpreted as the Boulware-Deser ghost and the DL provides a powerful tool to investigate the presence of this ghost in a given theory (as first argued in Refs. \cite{Deffayet:2005ys,Creminelli:2005qk}).  

\section{From DGP to dRGT gravity}
\label{NEWMASS}
\subsection{DGP gravity in brief}
\label{DGP}
The simplest DGP model \cite{Dvali:2000hr} is a five dimensional (5D in the following) brane-world model, and as such, it describes our four dimensional (4D in the following) Universe as a surface embedded into a 5D bulk space-time. Standard matter fields are thought as being localized on this surface, while the gravitational fields are living in the whole bulk space-time. The characteristic feature of the DGP model lies in the gravitational dynamics.  This dynamics is governed by the sum of two actions, the first one is a bulk gravitational action, which is the usual action for 5D gravity, namely
\begin{equation} \label{S5}
S_{(5)}=\frac{M_{(5)}^2}{2} \int d^5X \sqrt{g_{(5)}} R_{(5)},
\end{equation}
where   $R_{(5)}$ is the 5D Ricci scalar computed from the 5D metric $g^{(5)}_{ab}$ and $M_{(5)}$ is the reduced 5D Planck mass. To account for the brane, one adds to this action  a term of the form
\begin{equation} \label{S4}
S_{(4)}=\int d^4 x \sqrt{g_{(4)}} \left( \frac{M_P^2}{2} R^{(4)}+ {\cal L}_{(M)} \right),
\end{equation}
where ${\cal L}_{(M)}$ is a Lagrangian for brane localized matter (that is to say baryonic matter, dark matter, ...), and $R^{(4)}$ is the Ricci scalar of the  induced metric $g^{(4)}_{\mu \nu}$ on the brane. It is this term, depending on $R^{(4)}$, that is responsible for all the peculiarities of the gravitational phenomenology of DGP gravity.\footnote{This term is expected to be present in generic brane world constructions. It is the hierarchy of scales between $M_P$ and $M_{(5)}$ that is the real distinctive feature of DGP gravity.} The induced metric $g^{(4)}_{\mu \nu}$ is the metric experienced by the matter we are made of, it is defined by $
g^{(4)}_{\mu \nu} = \partial_\mu X^a \partial_\nu X^b g^{(5)}_{ab}, $ where $X^a(x^\mu)$ are defining the brane position in the bulk ($X^a$ being bulk coordinates  and  $x^\mu$ coordinates along the brane world-volume).

In this model, the gravitational potential between two static sources, separated by a distance $r$, interpolates between a 4D $1/r$ behavior at small distances and a 5D $1/r^2$ behavior at large distances, as shown in reference~\cite{Dvali:2000hr}. The crossover distance $r_c$ between the two regimes is given by 
\begin{equation} \label{TRANSRAD}
r_c \equiv  \frac{M^2_{P}}{2 M_{(5)}^3}.
\end{equation}
However, gravity does not reduce to a Newtonian potential, and in the DGP model, from a 4D point of view, gravity is mediated by a continuum of massive gravitons (so-called Kaluza-Klein modes), with no normalizable massless graviton entering into the spectrum.  As a consequence, the tensorial structure of the graviton propagator was shown to be the one of a massive graviton and the model shares some properties with the previously introduced NLFP theories. As we said, DGP gravity features in particular the vDVZ discontinuity.

A decoupling limit can be obtained in DGP gravity as well. In the scalar (i.e. helicity 0) sector, the obtained theory has a Lagrangian reading 
  \cite{Luty:2003vm,Nicolis:2004qq}
\begin{equation} \label{DLDGP}
 {\cal L}_{dgp} = 3 \tilde{\phi} \Box \tilde{\phi} -\frac{1}{\Lambda_{dgp}^3}(\de_{\mu} \tilde{\phi} )^2 \Box \tilde{\phi}
 + \frac{ \tilde{\phi} T}{2 M_p},
\end{equation}
where the strong coupling scale is given by $\Lambda_{dgp}=
 (M_p/r_c^2)^{1/3}$. Again the presence of the cubic operator in the above Lagrangian allows to understand easily the Vainshtein mechanism and associated scalings. Note however that the field equations deriving from this Lagrangian are just second order in line with the expectation that the DGP model does not contain any Boulware-Deser ghost  \cite{Deffayet:2005ys}. This remark is also at the root of the introduction of the Galileon family \cite{Nicolis:2008in}.

\subsection{dGRT gravity in brief}
\label{dRGT}
After the discovery of the BD ghost, the question arose if this ghost could be eliminated by a suitable choice of NLFP theory. This question was raised in the article of Boulware and Deser of 1972 \cite{Boulware:1973my}, and later, using the decoupling limit, in Ref. \cite{Creminelli:2005qk} which concluded that it was in fact not possible. The same question has been recently reexamined by de Rham, Gabadadze et Tolley \cite{deRham:2010kj,deRham:2010ik,deRham:2011rn}, who, using first the decoupling limit, built a family of theories (henceforth dRGT theories) which appeared (in the DL) to be devoid of the Boulware-Deser ghost. This was later confirmed by a  Hamiltonian analysis of the untruncated theory \cite{Hassan:2011hr,Hassan:2011ea} and several other works \cite{Kluson}, even though some authors still disagree \cite{Alberte:2010qb,Chamseddine:2011mu,Chamseddine:2013lid}.

dRGT theories are NLFP theories where the interaction term $S_{int}[f,g]$ of action (\ref{action}) has a special form that we will specify below in Eq.~(\ref{dGTBETA}). To do so, we use a parametrization given in references \cite{Hassan:2011hr,Hassan:2011vm}, we introduce functions  $\E_k$ defined for an arbitrary $n \times n$ matrix\footnote{With $I$, a line index belonging to $\{1,...,n\}$,  and $J$, a column index belonging to  $\{1,...,n\}$, and $n$ having so far no relation with the space-time dimension $D$.} $X^I_{\hphantom{I} J}$, represent elementary symmetric functions of the eigenvalues of $X$, and given, for $k=0$ to $4$, by (cf. e.g.  \cite{Hassan:2011hr}) 
\begin{equation}\label{eX}
\eqalign{ 
e_0 \left(X\right) &= 1,\;
e_1 \left(X\right) = [X] ,\;
e_2 \left(X\right) = \frac{1}{2} \left([X]^2 - [X^2]\right)\\
e_3 \left(X\right) &= \frac{1}{6} \left([X]^3 - 3 [X][X^2] + 2 [X^3]\right) \\
e_4 \left(X\right) &= \frac{1}{24} \left([X]^4 - 6[X]^2[X^2] +3[X^2]^2 + 8 [X][X^3]- 6 [X^4]\right)}
\end{equation}
where $[X]$ designate the trace  $X^I_{\hphantom{I} I}$ of the matrix $X$. For an arbitrary integer $k$, one defines $\E_k$ by 
\be \label{defEk}
\E_k(X) = \frac{1}{k!} X^{I_1}{}_{[I_1}...X^{I_k}{}_{I_k]}, 
\ee
where brackets $[\;]$ around indices indicate the unnormalized antisymmetric sum over permutations. For a matrix $X$,  Cayley-Hamilton theorem implies  that  
\be \label{defEn}
{\rm det}  (X) = \E_n(X). 
\ee
dRGT theories \cite{deRham:2010kj,deRham:2010ik,deRham:2011rn} can be defined by an action of the form  \cite{Hassan:2011hr} 
\be \label{dGTBETA}
S = M_P^2 \int d^4 x \sqrt{-g}\left[R -  m^2 \sum_{k=0}^{k=4} \beta_k \E_k\left(\sqrt{g^{-1} f}\right)\right]
\ee
where $\beta_n$ are arbitrary parameters, and the square root above is a matrix square root of the tensor  ${\bf g}^{-1}{\bf f}$\footnote{Note that this square root does not always exist. However, a vierbein formulation (such as given in \cite{Hinterbichler:2012cn}) of the theory allows to show that the existence of the square root does follow from the field equations, at least in a subset of theories. It also provides a simple way to count the number of degrees of freedom \cite{Hinterbichler:2012cn,US}.}. One can notice, on the one hand that $\beta_0$ describes a cosmological constant (which does not give any mass to the graviton), on the other hand that the term proportional to $\beta_4$ does not give any contribution to the field equations of  $g_{\mu \nu}$, since  $\sqrt{-g}\; \E_4\left(\sqrt{g^{-1} f}\right) = \sqrt{-g} \;{\rm det} \left(\sqrt{g^{-1} f}\right) = \sqrt{-f}$. Hence, for 
$D=4$ dimensions,  we get a three parameter family of massive theories parametrized by  $\beta_k$, with  $k=1,2,3$ (which becomes a two parameter family once the mass of the graviton is fixed). It is easy to extend this construction to  $D$ dimensions by considering actions of the form 
\be \label{dGTBETAd}
S = M_P^{D-2} \int d^D x \sqrt{-g}\left[R - 2\Lambda -  m^2 \sum_{k=1}^{k=D-1} \beta_k \E_k\left(\sqrt{g^{-1} f}\right)\right]
\ee
where  $\E_k$ are defined as in (\ref{defEk}). The interaction term $S_{int}[f,g]$ of theories (\ref{dGTBETA})-(\ref{dGTBETAd}) is indeed of the form (\ref{INT}). 

\subsection{Decoupling limit of dRGT model}
A decoupling limit can also be obtained in the dRGT theory, as we now recall.
It will be more convenient for that purpose to use an alternative form for the action (\ref{dGTBETA}), written in terms of the matrix $\mathcal{K}$,
\begin{equation}
	\mathcal{K} = {\mathbb I} - \sqrt{\bf g^{-1} f}.
\end{equation}
The dRGT action then reads,
\begin{eqnarray}
\label{dGTBETAa}
S = M_P^2 \int d^4 x \sqrt{-g}\left[R + 2 m^2 \left(e_2\left(\mathcal{K}\right)+\alpha_3e_3\left(\mathcal{K}\right) +\alpha_4e_4\left(\mathcal{K}\right) \right)\right],
\end{eqnarray}
with the following identifications, 
\begin{eqnarray}
\beta_0 = -12  -8\alpha_3 -2 \alpha_4,\, \beta_1 = 6 + 6\alpha_3+ 2\alpha_4,\nonumber\\
 \beta_2 = -2 -4\alpha_3-2\alpha_4,\, \beta_3 = 2(\alpha_3+\alpha_4)\nonumber
\end{eqnarray}
As in the case of NLFP, we introduce the ``Goldstone boson''  expansion, defined by (\ref{STUCA}), (\ref{DECPIPI}) and (\ref{DECPIbis}). The matrix $\mathcal{K}$ defining the interaction term has then the form,
\begin{equation}
\label{K1}
\mathcal{K}^\mu_\nu =\delta^\mu_\nu- \sqrt{\delta^\mu_\nu- g^{\mu\alpha}H_{\alpha\nu}},
\end{equation}
where $H_{\mu\nu}$ is given by (\ref{EXPANDH}).
Expansion of (\ref{dGTBETAa}) in powers of $\hat H\equiv H^\mu_\nu = (g^{-1} H)$ gives,
\begin{eqnarray}
\label{ExpandinH}
\eqalign{
e_2&\left(\mathcal{K}\right)+\alpha_3e_3\left(\mathcal{K}\right) +\alpha_4e_4\left(\mathcal{K}\right) =  \frac18\left([\hat{H}]^2-[\hat{H}^2]\right) \\
	 +& \frac{1}{16}\left(\frac{\alpha_3}{3}[\hat{H}]^3+(1-\alpha_3)[\hat{H}][\hat{H}^2]-(1-\frac{2\alpha_3}{3})[\hat{H}^3]\right) \\
	+ &\frac{1}{128}\Big\{ \frac{\alpha_4}{3}[\hat{H}]^4 + 2(\alpha_3-\alpha_4)[\hat{H}]^2[\hat{H}^2] +(1-2\alpha_3+\alpha_4)[\hat{H}^2]^2 \\
	+& 4\left(1-\alpha_3+\frac{2\alpha_4}{3}\right)[\hat{H}][\hat{H}^3] -\left(5-4\alpha_3+2\alpha_4\right)[\hat{H}^4]\Big\} + \mathcal{O}(\hat H^5).\\
	}
\end{eqnarray}
Up to redefinition of constants, (\ref{ExpandinH}) coincides with equations obtained in \cite{deRham:2010ik} and \cite{Koyama:2011yg}. 
Substituting (\ref{ExpandinH}) with $\hat{H}$ expressed as (\ref{EXPANDH}) into (\ref{dGTBETAa}), 
one obtains the action as a series expansion in  $h_{\mu\nu}$, $A_\mu$ and $\phi$. 
Because of the particular form of $e_2(\mathcal{K})$ in (\ref{dGTBETAa}), chosen to recover the Fierz-Pauli theory,
the quadratic part of the action leads to the same expression, as in general NLFP, Eq.~(\ref{qexpbis}) 
(up to an overall factor 2, due to different normalizations in the actions).
As we already mentioned before, it is possible to arrange the interaction term of a generic NLFP to  
remove the leading interactions $\sim (\partial\tilde\phi)^3/(M_Pm^4)$. 
In fact, the dRGT theory does exactly this: the interactions $\sim (\partial\tilde\phi)^3/(M_Pm^4)$ cancel 
and the leading interaction is then suppressed by a higher scale\footnote{Note that the cubic vector-scalar terms $\sim \partial A\partial ^2\tilde\phi \partial ^2\tilde\phi$ are suppressed by $\Lambda_4\ll \Lambda_3$, however, these terms cancel out up to a total derivative \cite{deRham:2010gu}.}
\begin{equation}
\Lambda_3 = \left(m^2 M_P\right)^{1/3}.
\end{equation}
Therefore for the dRGT model we define the decoupling limit as 
\begin{equation}
\label{L3DL}
M_P \to \infty,\, m \to 0, \, \Lambda_3 \sim {\rm const}, \; T_{\mu\nu}/M_P\sim {\rm const}.
\end{equation}
Since the vector mode does not couple to a source, we can set it consistently to zero, $A_\mu =0$. 
Then in the decoupling limit (\ref{L3DL}) we  obtain,
\begin{equation}
\label{DLdRGT}
\eqalign{
S =&\int d^{4}x
\Big\{ -\frac12\hat{h}^{\mu\nu}\mathcal{E}^{\alpha\beta}_{\mu\nu}\hat{h}_{\alpha\beta}\\
&+\hat{h}^{\mu\nu} X^{(1)}_{\mu\nu} +\frac{\tilde{\alpha}}{\Lambda^3_3} \hat{h}^{\mu\nu} X^{(2)}_{\mu\nu}  
	+\frac{\tilde{\beta}}{\Lambda_3^6} \hat{h}^{\mu\nu} X^{(3)}_{\mu\nu} 
+T_{\mu\nu}\hat{h}^{\mu\nu}\Big\},
}
\end{equation}
where  
the following notations are introduced:
$\tilde\alpha = 1+\alpha_3$, $\tilde\beta = \alpha_3 + \alpha_4$,
\bea
X^{(1)}_{\mu\nu} &=& \frac12 \epsilon_\mu^{\phantom{\mu}\alpha\rho\sigma}\epsilon_{\nu\phantom{\beta}\rho\sigma}^{\phantom{\nu}\beta}\Phi_{\alpha\beta}, \label{X1mn}\\
 X^{(2)}_{\mu\nu} &=& -\frac12 \label{X2mn} \epsilon_\mu^{\phantom{\mu}\alpha\rho\gamma}\epsilon_{\nu\phantom{\beta\sigma}\gamma}^{\phantom{\nu}\beta\sigma}\Phi_{\alpha\beta}\Phi_{\rho\sigma},\\
 X^{(3)}_{\mu\nu} &=& \frac16 \label{X3mn} \epsilon_\mu^{\phantom{\mu}\alpha\rho\gamma}\epsilon_{\nu}^{\phantom{\nu}\beta\sigma\delta}\Phi_{\alpha\beta}\Phi_{\rho\sigma}\Phi_{\gamma\delta},
\eea
with  $\Phi_{\mu\nu}=\tilde\phi_{,\mu\nu}$,  and $\epsilon$ the  totally antisymmetric Levi-Civita tensor. 
Note that as it was argued in \cite{deRham:2010ik,deRham:2010kj}
the higher order terms in the decoupling limit  are equal to zero, therefore we stopped the expansion in  (\ref{ExpandinH}) at the fourth order. 
The equations of motion obtained from (\ref{DLdRGT}) read,
\begin{eqnarray}
\mathcal{E}^{\alpha\beta}_{\mu\nu}\hat{h}_{\alpha\beta} 
- X^{(1)}_{\mu\nu} - \frac{\tilde\alpha}{\Lambda^3_3} X^{(2)}_{\mu\nu}  - \frac{\tilde\beta}{\Lambda_3^6} X^{(3)}_{\mu\nu}  = \frac{T_{\mu\nu}}{M_P},\label{EOMdRGTh0}\\
\partial_\alpha\partial_\beta \hat{h}^{\mu\nu}\left(
\frac12\epsilon_\mu^{\phantom{\mu}\alpha\rho\sigma}\epsilon_{\nu\phantom{\beta}\rho\sigma}^{\phantom{\nu}\beta}
-\frac{\tilde\alpha}{\Lambda^3_3} \epsilon_\mu^{\phantom{\mu}\alpha\rho\gamma}\epsilon_{\nu\phantom{\beta\sigma}\gamma}^{\phantom{\nu}\beta\sigma}\Phi_{\rho\sigma}
+\frac{\tilde\beta}{2\Lambda^6_3}\epsilon_\mu^{\phantom{\mu}\alpha\rho\gamma}\epsilon_{\nu}^{\phantom{\nu}\beta\sigma\delta}\Phi_{\rho\sigma}\Phi_{\gamma\delta}
\right) =0 \nonumber \label{EOMdRGTp0}\\
\end{eqnarray}
Notice that the helicity-0 and helicity-2 modes are mixed at  all orders. 
It is possible, however, to diagonalize the linear and quadratic mixings. 
Indeed, making here the nonlinear field redefinition in (\ref{DLdRGT}) \cite{deRham:2010ik} (as opposed to the linear redefinition (\ref{defhhatht}) used before),
\begin{equation}
\label{tildehhath}
\hat h_{\mu\nu}= \tilde{h}_{\mu\nu} - \eta_{\mu\nu}\tilde\phi - \tilde\alpha\frac{\partial_\mu\tilde{\phi}\partial_\nu\tilde{\phi}}{\Lambda_3^3},
\end{equation}
and using the relations $\mathcal{E}^{\alpha\beta}_{\mu\nu}(\eta_{\alpha\beta}\tilde{\phi}) = -X^{(1)}_{\mu\nu}$ and 
$\mathcal{E}^{\alpha\beta}_{\mu\nu}(\partial_\alpha\tilde{\phi}\partial_\beta\tilde{\phi}) = -X^{(2)}_{\mu\nu}$ one obtains up to total derivatives,
\begin{equation}
\label{DLdRGT1}
\eqalign{
S =&\int d^{4}x
\Big\{ -\frac12\tilde{h}^{\mu\nu}\mathcal{E}^{\alpha\beta}_{\mu\nu}\tilde{h}_{\alpha\beta}
+ \frac32\tilde\phi\Box\tilde\phi 
-\frac{\tilde\alpha}{\Lambda_3^3}\tilde{\phi}^{,\mu}\tilde{\phi}^{,\nu}X^{(1)}_{\mu\nu}\\
&
-\frac{1}{\Lambda_3^6}\left(\frac{\tilde\alpha^2}{2}+\frac{\tilde\beta}{3}\right) \tilde{\phi}^{,\mu}\tilde{\phi}^{,\nu}X^{(2)}_{\mu\nu} 
+\frac{\tilde\beta}{\Lambda_3^6} \left( h^{\mu\nu} -\frac{\tilde\alpha}{\Lambda_3^3}\tilde{\phi}^{,\mu}\tilde{\phi}^{,\nu}\right)X^{(3)}_{\mu\nu} \\
& +\frac{1}{M_P}\left(T_{\mu\nu}\tilde{h}^{\mu\nu} -T\tilde{\phi} -\frac{\tilde\alpha}{\Lambda_3^3}T_{\mu\nu}\partial^\mu\tilde{\phi}\partial^\nu\tilde{\phi}\right)
\Big\}.
}
\end{equation}
One can recognize the Galileon \cite{Nicolis:2008in,Deffayet:2009wt} kinetic terms in (\ref{DLdRGT1}). 
Notice that the kinetic mixing does not disappear in the general case $\tilde\beta \neq 0$. Only in the special case $\tilde\beta =0$, 
the helicity-0 and helicity-2 modes are completely decoupled. 
It is also worth to note that $\tilde{\phi}$ couples directly to matter source in a ``disformal'' way.
Varying (\ref{DLdRGT1}) with respect to $\tilde h_{\mu\nu}$ and $\tilde{\phi}$ respectively, we obtain,
\begin{eqnarray}
\mathcal{E}^{\alpha\beta}_{\mu\nu}\tilde{h}_{\alpha\beta} & -& \frac{\tilde\beta}{\Lambda_3^6} X^{(3)}_{\mu\nu}  = \frac{T_{\mu\nu}}{M_P},\label{EOMdRGTh}\\
3\Box\tilde\phi &+& \frac{3\tilde\alpha}{\Lambda_3^3}\Phi^{\mu\nu} X^{(1)}_{\mu\nu}
+\frac{4}{\Lambda_3^6}\left(\frac{\tilde\alpha^2}{2}+\frac{\tilde\beta}{3}\right)\Phi^{\mu\nu} X^{(2)}_{\mu\nu} 
+\frac{5\alpha\beta}{\Lambda^3_3}\Phi^{\mu\nu} X^{(3)}_{\mu\nu}  \nonumber\\
&+&\frac{\tilde\beta}{2\Lambda^6_3}\partial_\alpha\partial_\beta \tilde{h}^{\mu\nu}\epsilon_\mu^{\phantom{\mu}\alpha\rho\gamma}\epsilon_{\nu}^{\phantom{\nu}\beta\sigma\delta}\Phi_{\rho\sigma}\Phi_{\gamma\delta}
= \frac{T}{M_P} - \frac{2\tilde\alpha}{M_P\Lambda^3_3}\Phi^{\mu\nu}T_{\mu\nu}.\label{EOMdRGTp}
\end{eqnarray}
The above equations describe a GR-like graviton $\tilde{h}_{\alpha\beta}$ and the helicity-0 piece of the massive graviton $\tilde{\phi}$.
The helicity-1 modes drops out from the resulting equations because of the use of DL. 
Note that $\tilde{h}_{\alpha\beta}$ is not ``physical'', it is connected to the one measured in experiments, $\hat h_{\mu\nu}$, by Eq.~(\ref{tildehhath}).
The nonlinear structure of~(\ref{EOMdRGTp}) 
is responsible for the GR restoration via the Vainshtein mechanism as we will show in the next section.

\section{The Vainshtein mechanism}
\label{Vain}
\subsection{The Vainshtein mechanism in the decoupling limits and k-mouflage}
The Vainshtein mechanism  was introduced by Vainshtein by looking at static spherically symmetric ans\"atze in a theory close to one belonging to the NLFP family discussed before. Such a theory, at large distances from a source, can be described linearly, so that all modes of the massive graviton (i.e. including the ones additional to the two polarizations of a massless graviton) propagate, resulting in large deviations from GR responsible for the vDVZ discontinuity. 
Vainshtein pointed out that, at some distance to the source, called today the Vainshtein radius, $r_V$, the linear regime breaks down, and for $r$ below $r_V$ the theory enters a nonlinear regime. The Vainshtein radius is a composite scale made out of the Planck mass, a scale specific to the theory considered (e.g. the graviton mass) and the mass of the source. Vainshtein also proposed to find the solution below the Vainshtein radius using an expansion in terms of the mass of the graviton, instead of that in terms of  the Newton's constant that is used at large distance. However he did not show that the two expansions could be matched in an existing solution. That such a solution exists was only proven recently in Refs. \cite{Babichev:2009us,Babichev:2009jt,Babichev:2010jd}. However, the Vainshtein mechanism was re-introduced before, in the framework of DGP gravity \cite{Deffayet:2001uk}. There, indeed, exact solutions - however of cosmological type - were found \cite{Deffayet:2000uy,Deffayet:2001aw} featuring for the first time an interpolation between two regimes: one exhibiting properties similar to those of the linearized regime of NLFP theories and another recovering GR\footnote{Note that exact (and proven to be everywhere non singular) solutions appropriate to describe a static spherically symmetric space-time on the brane in the DGP model, which would be analogous there to the solutions found in \cite{Babichev:2009us,Babichev:2009jt,Babichev:2010jd}, are not known, despite a lot of effort to find them \cite{Gruzinov:2001hp,Porrati:2002cp,Lue:2002sw,Gabadadze:2004iy}.}.

The Vainshtein mechanism was then used and believed (and in some cases has been shown) to allow the recovery of General Relativity around massive bodies, and more generally in cases where the space-time curvature is larger than some critical value, for a wide class of non-GR theories of gravity. Roughly speaking, this is achieved by hiding extra degree(s) of freedom by strong kinetic self-coupling, so that they almost do not propagate. That it can work can already be seen in the decoupling limit, described in the previous sections, which allows to separate  the different helicity modes and to focus on the features crucial for the Vainshtein mechanism.
To introduce with some details this mechanism, and illustrate it in action, we start then to discuss it within the framework of this decoupling limit, which is much simpler. This framework can also be obtained without referring to massive gravity from k-mouflage models \cite{Babichev:2009ee}, 
 generic scalar-tensor theories with kinetic self-interactions in the scalar sector, which share many features with massive gravity theories, as we will see below.

\subsubsection{Vainshtein mechanism and k-mouflage.} 
\label{KMOUFLAGE}
Our starting point will be here the action
\begin{equation}
\label{ACTIONkm1}
\eqalign{
S =&\int d^{4}x
\Big\{ -\frac12\hat{h}^{\mu\nu}\mathcal{E}^{\alpha\beta}_{\mu\nu}\hat{h}_{\alpha\beta}
+\hat{h}^{\mu\nu} X^{(1)}_{\mu\nu}  + M_P^2 m^2 K_{NL} +T_{\mu\nu}\hat{h}^{\mu\nu}\Big\}.
}
\end{equation}
which is a quadratic action for the (normalized) graviton $\hat{h}_{\mu \nu}$, and describes in addition a scalar degree of freedom $\tilde\phi$ (which has here the dimension of a mass) with a non linear action density $K_{NL}$ which is a dimensionless nonlinear function of $\tilde\phi$ and its derivatives (not necessary only first derivatives).
Note that $\tilde\phi$ has a kinetic mixing with $\hat{h}_{\mu \nu}$ through the second term on the right hand side above. 

This action can be obtained from the following covariant action 
\begin{equation}
\label{ACTIONkm}
S_{k-mouflage} = M_P^2\int d^4 x \sqrt{-g}\left[ R + \frac{\tilde\phi}{2M_P}  R + m^2 K_{NL} \right] + S_m[g],
\end{equation}
which defines the k-mouflage family \cite{Babichev:2009ee}. In the regime of weak gravity we can expand (\ref{ACTIONkm}) in perturbations (in terms of the unnormalized graviton $h_{\mu \nu}$) around Minkowski space-time. If we do so and keep the Einstein-Hilbert term up to $h^2$, keep only the first order in $h$ in the mixing term $\sim\tilde\phi R$ (this because $\tilde\phi$ will be of order of $\hat h$  or  of higher order as we will verify later), and do not expand $K_{NL}$ (because the Vainshtein mechanism relies precisely on  the non-linearity of this term), we obtain, after normalizing the spin-2 perturbations (as $h_{\mu\nu}\to \hat{h}_{\mu\nu}/M_P$), precisely the action (\ref{ACTIONkm1}), up to a total derivative.

Coming back to action (\ref{ACTIONkm1}) and redefining the spin-2 mode as $\hat h_{\mu\nu}= \tilde{h}_{\mu\nu} - \eta_{\mu\nu}\tilde\phi$, we get
\begin{equation}
\label{ACTIONkm2}
S =\int d^{4}x
\Big\{ -\frac12\tilde{h}^{\mu\nu}\mathcal{E}^{\alpha\beta}_{\mu\nu}\tilde{h}_{\alpha\beta}
+ \frac32\tilde\phi\Box\tilde\phi + M_P^2 m^2K_{NL}
+\frac{1}{M_P}\left(T_{\mu\nu}\tilde{h}^{\mu\nu} -T\tilde{\phi} \right)
\Big\}.
\end{equation}
Note that similar to the decoupling limit of massive gravity, the spin-0 and spin-2 modes are decoupled, and a non-minimal scalar-matter coupling appears. The equations of motion following from (\ref{ACTIONkm2}) read,
\begin{eqnarray}
\mathcal{E}^{\alpha\beta}_{\mu\nu}\tilde{h}_{\alpha\beta}   = \frac{T_{\mu\nu}}{M_P},\label{EOMkmh}\\
3\Box\tilde\phi + \mathcal{E}_\phi = \frac{T}{M_P},\label{EOMkmp}
\end{eqnarray}
where $\mathcal{E}_\phi \equiv M_P^2 m^2 \delta K_{NL}/\delta \tilde\phi$.
The essence of the Vainshtein mechanism can be easily seen from the two equations above. 
First, noticing that when the linear term is dominant in the l.h.s. of (\ref{EOMkmp}), the solution for $\tilde\phi$ is of order of $\tilde{h}_{\mu\nu}$, 
obtained from (\ref{EOMkmh}). Hence the (normalized) physical metric $\hat h\sim \tilde{h} + \tilde\phi$ (see Eq.~(\ref{defhhatht})) therefore receives $\mathcal{O}(1)$ corrections as in a free scalar tensor theory. On the other hand,  there is usually a regime where the nonlinear term $\mathcal{E}_\phi$ is dominant in (\ref{EOMkmp}), then $\tilde\phi$ is subdominant 
in comparison to $\tilde{h}$, therefore $\hat h\simeq \tilde{h}$ and GR is restored.

The detailed study of the k-mouflage model for various nonlinear terms $K_{NL}$ in the static and spherically symmetric case confirms the arguments above \cite{Babichev:2009ee}. In this case, as well as for what follows, it is convenient to choose the Newtonian gauge 
(see e.g. Ref.~\cite{Deffayet:2004ru} where the vDVZ discontinuity, in this gauge, was studied in the context of the DGP model, 
and a more recent work~\cite{Alberte:2010it} for massive gravity) for the metric written in the spherical coordinates,
\begin{equation}
\label{NG}
	ds^2 = -(1+\Psi/M_P)dt^2 + (1-\Phi/M_P)(dr^2 + r^2d\Omega^2).
\end{equation}  
The independent components of the linearized Einstein tensor in the gauge (\ref{NG}) read,
$ \mathcal{E}^{\alpha\beta}_{00}\hat h_{\alpha\beta} =\frac{1}{r^2} \left(r^2\Phi'\right)'$,
$\mathcal{E}^{\alpha\beta}_{01}\hat h_{\alpha\beta} =\frac{1}{r} \left(\Psi-\Phi\right)'$.
The linearized GR solution is then $\Psi_{GR} = \Phi_{GR} =-M_P r_S/r$. Typically, 
the scale where the nonlinear regime switches on depends on the form of $\mathcal{E}_\phi$. 
We can parametrize the nonlinear term as $\mathcal{E}_\phi \sim \partial^{n-k+3}\tilde\phi^k/ \Lambda^{n}_n$, with 
$\Lambda_n^n = M_P m^{n-1}$. The Vainshtein radius --- where the regime changes from linear to nonlinear --- 
can be found by comparing the two terms in the l.h.s. of (\ref{EOMkmp}) and using $\partial \to 1/r$, ($r$ being the distance to the source). One finds \cite{Babichev:2009ee}
\begin{equation}
\label{GENrV}
r_V = \frac{1}{\Lambda_n} \left(M_P r_S\right)^{(k-1)/n}.
\end{equation}
Note that the galileon model \cite{Nicolis:2008in,Deffayet:2009wt} is precisely of the k-mouflage type, therefore one naturally expects 
the Vainshtein mechanism to be operating in this model as well. Indeed, the study of static spherically symmetric configurations confirms this \cite{Nicolis:2008in,DeFelice:2011th}.

It should be stressed, however, that it is not enough to establish the existence of the two regimes: the Vainshtein one, where GR is restored, and the linear one, far from the source, where the solution for the linearized theory (e.g. the Fierz-Pauli linear massive gravity) is recovered. It should be also checked that there exists an everywhere non singular solution that matches the two regimes. This crucial step is lacking in the original paper of Vainshtein as was stressed in particular in Ref. \cite{Boulware:1973my} and, it is only recently that the existence of such a solution was shown \cite{Babichev:2009us,Babichev:2009jt,Babichev:2010jd} after some previous contradictory claims \cite{Jun:1986hg,Damour:2002gp}. 
We will come back below on these results. Let us before introduce the Vainshtein mechanism successively in the decoupling limits of generic NLFP and dRGT theories.

\subsubsection{Massive gravity with general potential.} In this case, looking only at the decoupling limit, one notices that the equations of motion for spin-2 and spin-0 have the form identical to those of the decoupling limit of k-mouflage (\ref{ACTIONkm1}).
Indeed,  equation (\ref{tildephi}) coincides with (\ref{EOMkmp}) with $\mathcal{E}_\phi \sim \partial^{6}\tilde{\phi}^2/ \Lambda^{5}_5$.
One then expects, from the arguments above, that GR is restored inside $r_V = (r_S/m^4)^{1/5}$.
Integrating (\ref{tildephi}) once yields a second order nonlinear differential equation,
\begin{equation}
\frac{2}{\Lambda^{5}_5}Q(\tilde{\mu})+\frac{3}{2}\;\tilde{\mu}=\frac{M_P r_S}{r^{3}}, \label{FIbis}
\end{equation}
where we have defined $\tilde{\mu}$ as 
\beq \tilde{\mu}=-\frac{2}{r}\tilde{\phi}', \label{defmuphit}
\eeq 
and $Q(\tilde{\mu})$ is a non-linear second-order function of $\tilde{\mu}$, 
containing terms $\tilde{\mu}\tilde{\mu}'$, $\tilde{\mu}'^2$, $\tilde{\mu}\tilde{\mu}''$ and $\tilde{\mu}'\tilde{\mu}''$. 
The function $Q(\tilde{\mu})$ is in one to one correspondence with the quadratic terms in the right hand side of Eq.(\ref{tildephi}) (for details see \cite{Babichev:2009us}). An integration constant has been chosen so as to match the source, resulting in the r.h.s. of (\ref{FIbis}). The quantity $\tilde{\mu}$ has here the meaning of a gauge function that goes from a unitary gauge where the flat (fiducial) metric $f_{\mu \nu}$ has just the  Minkowski form in spherical coordinates to a non unitary gauge where the only non trivial components of the physical metric are $g_{tt}$ and $g_{rr}$ (like in the Schwarzschild form)} while the fiducial flat metric contains $\tilde{\mu}$ \cite{Babichev:2009us}. 
Depending on the form of $Q(\tilde\mu)$ (that is, on the form of the potential term), a global solution matching the flat asymptotic behavior and the Vainshtein regime  may or may not exist. 
In particular, for the potential (\ref{S2}) there is no such solution. 
On the other hand, for the potential (\ref{S3}), such a solution exists. 
Let us study this case in more detail as an illustrative example.
For the potential (\ref{S3}) we have 
\begin{equation}
Q(\tilde{\mu}) = -\frac{\tilde\mu'^2}{4} - \frac{\tilde\mu\tilde\mu''}{2} - \frac{2\tilde\mu\tilde\mu'}{r}.\label{QDL}
\end{equation}
In the linear regime, (\ref{FIbis}), (\ref{defmuphit}) and (\ref{EOMkmh}) give,
\begin{equation} \label{GNexp}
\eqalign{
\Psi &= -\frac{4M_P}3\frac{r_S}{r}\left[1 + \mathcal{O}\left(\frac{r_V}{r}\right)^5\right], \, \Phi = -\frac{2M_P}3\frac{r_S}{r}\left[1 + \mathcal{O}\left(\frac{r_V}{r}\right)^5\right], \\
\tilde{\phi} & = \frac{M_P}3\frac{r_S}{r}\left[1 + \mathcal{O}\left(\frac{r_V}{r}\right)^5\right],
}
\end{equation}
As it was expected, deviations from GR are of order of one in this regime. In particular ones notes that, in this regime, $\Psi = 2 \Phi$ which is another way to express the vDVZ discontinuity. Note also that we recognize above the first terms of an expansion in the Newton constant (or equivalently in the Schwarzschild radius of the source). The expansion  breaks down at the Vainshtein radius where the non linear regimes switches on and $  \mathcal{O}\left(\frac{r_V}{r}\right) \sim 1$. In this second regime, we obtain from (\ref{FIbis}),
\begin{equation}
	\label{GNV}
	\Psi = \Phi = - \frac{M_P r_S}{r} + \frac{M_P r_S}{r}\times \mathcal{O}\left(\frac{r}{r_V}\right)^{5/2},\,
	\tilde{\phi} = -\frac{2\sqrt{2}M_P}{9} \frac{r_S}{r}\left(\frac{r}{r_V}\right)^{5/2}.
\end{equation}
Therefore GR is restored for $r\ll r_V$ as it was anticipated, since we now have  $\Psi = \Phi$ at dominant order. Moreover, as announced above, the corrections to GR (as well as $\tilde \phi$) are seen to be $\propto r_V^{-5/2}$ which is found to be proportional to the square of the mass of the graviton $m^2$. These corrections represent the first terms of an expansion around GR solutions in terms of $m^2$. The fact that the two regimes, Eq.~(\ref{GNexp}) and Eq.~(\ref{GNV}) match each others in an existing solution can be checked by solving numerically 
Eq.~(\ref{FIbis}), as first shown in Ref. \cite{Babichev:2009us}. This latter reference also gives a rigorous study of the Vainshtein mechanism for nonlinear massive gravity in the decoupling limit 
for general potentials (with the only restriction that such potentials lead to non vanishing cubic operators suppressed by $\Lambda_5$ in the DL, which for example does not apply for the dRGT model). Note also that another derivation of (\ref{GNV}) using a slightly different approach has been presented in~\cite{Alberte:2010it}.

For some choice of the potential in NLFP theory, there are solutions inside $r_V$, which do not feature the Vainshtein scaling, 
however, GR is also restored. This type of non-Vainshtein scaling is obtained by expanding (\ref{FIbis}) around the 
zero-mode of the operator $Q$ (\ref{QDL}). These solutions turns out to be unphysical, and it is interesting to compare this 
to the case of dRGT theory, where solutions other than Vainshtein-like also exist.

\subsubsection{dRGT massive gravity}  As in the case of  NLFP with a general potential term, the decoupling limit of
the dRGT model with the choice $\tilde\beta =0 $ also falls into the k-mouflage category. 
Indeed, the equations of motion for k-mouflage (\ref{EOMkmh})-(\ref{EOMkmp})  
reproduce those for the decoupling limit of dRGT model  (with $\tilde\beta=0$) (\ref{EOMdRGTh})-(\ref{EOMdRGTp}) with $\mathcal{E}_\phi$ of the Galileon type. 
For spherically symmetric static ansatz, (\ref{EOMdRGTp}) reads, in terms of $\tilde{\mu}$ defined as in (\ref{defmuphit})
\begin{equation}
	\frac32\tilde\mu +\frac{3\tilde\alpha}{2\Lambda_3^3} \tilde\mu^2 +\frac{\tilde{\alpha}^2}{4\Lambda_3^6}\tilde\mu^3 = \frac{M_P r_S}{r^{3}}.\label{Emu}
\end{equation}
The last equation, describing the scalar mode of the dRGT model is quite similar to that for massive gravity with a general potential (\ref{FIbis}),
with, however, an important difference: (\ref{Emu}) is algebraic, while (\ref{FIbis}) is a differential equation of the second order.
This can be traced back to the fact
that massive gravity with a general potential possesses an extra (BD-ghost) scalar degree of freedom compared to the dRGT model, and that, as we said before, the presence of this ghost manifests itself in the higher derivative nature of the field equations in the scalar sector, here given by those (as seen in the definition (\ref{defmuphit})) for $\tilde{\mu}$. Note also that in DL of DGP the cubic term in l.h.s. of (\ref{Emu}) is absent, resulting in different corrections to the Newtonian potential.

The Vainshtein radius for the dRGT model, indicating where the linear regime for this model breaks down, can be read off from (\ref{GENrV}) and reads
\beq  
r_V = \left(r_S/m^2\right)^{1/3}. \label{dRGTVain}
\eeq
In fact, since (\ref{Emu}) is an algebraic third order equation, one can write down explicitly three different branches  each corresponding to a branch of solutions. 
We will not need here all these branches, but only note that the Vainshtein regime is found from (\ref{Emu}) assuming the last term in the r.h.s. is dominant \cite{Koyama:2011xz},
\begin{equation}\label{dRGTVAINSHTEIN}
\eqalign{	\Psi  = - \frac{M_P r_S}{r} + \frac{M_P r_S}{r}\times \mathcal{O}\left(\frac{r}{r_V}\right)^{2},\,  \Phi = - \frac{M_P r_S}{r} + \frac{M_P r_S}{r}\times \mathcal{O}\left(\frac{r}{r_V}\right),\\
	\tilde{\phi} = -\frac{M_Pr_S}{(2\tilde\alpha^2)^{1/3}r} \left(\frac{r}{r_V}\right)^2.}
\end{equation}
Note that as in the case of NLFP with general potential, not  all $\tilde\alpha$ lead to a global solution 
 matching flat asymptotic at $r>r_V$ and  the Vainshtein regime at $r<r_V$. 
For $\tilde\alpha>0$ such a solution exists, and for $\tilde\alpha<0$ it does 
not~\cite{Nicolis:2008in,Koyama:2011xz
}\footnote{Different conditions for existence and stability however apply when there is cosmological evolution for the Galileon~\cite{Babichev:2012re}, in particular, an unstable model in asymptotically Minkowski space-time may be stable in the asymptotically de-Sitter. In the context of dRGT gravity it was argued~\cite{Berezhiani:2013dw} (see also a more recent paper~\cite{Koyama:2013paa}) that the choice $\tilde\alpha>0$ in fact is not physical (be aware of the difference in notation, 
$\tilde\alpha$ in our paper equals to $-\alpha$ of Ref.~\cite{Berezhiani:2013dw}), since the renormalization of the kinetic term for $\tilde\phi$ in e.g. (\ref{DLdRGT1}) is negative in the presence of a source due to the disformal coupling.
For $\tilde\alpha<0$ on the other hand there is a stable solution with a cosmological asymptotic.}.

 When $\tilde\beta\neq 0$, the decoupling limit of dRGT theory (\ref{DLdRGT}) does not exactly fall into the k-mouflage category. 
 In this case, indeed, it seems that no local redefinition of fields can decouple spin-0 and spin-2 modes~\cite{deRham:2010ik}. 
However, it is not difficult to refine the argument we used for the k-mouflage model. 
Indeed, anticipating the Vainshtein regime, we neglect the second term on the l.h.s. of (\ref{EOMdRGTh}) and
all the terms containing only scalar in (\ref{EOMdRGTp}), since we expect $\tilde h\gg \tilde\phi$.
To cancel the r.h.s. of (\ref{EOMdRGTp}), one should assume 
$\tilde\phi \sim r^2 M_P m^2/\sqrt{\tilde\beta} $, so that $\tilde\phi \ll \Phi$, hence GR is restored. 
More precisely, taking $\Psi = \Phi = \Phi_{GR}$ and substituting this into (\ref{EOMdRGTp}) and neglecting 
all the terms on r.h.s. but the mixing term, we find for the the spherically symmetric ansatz~\cite{Chkareuli:2011te}, 
\begin{equation}
	\Psi = \Phi = - \frac{M_P r_S}{r} + \frac{M_P r_S}{r}\times \mathcal{O}\left(\frac{r}{r_V}\right)^{3},\,
	\tilde{\phi} = -\frac{M_Pm^2}{\sqrt{\tilde\beta}} r^2,
\end{equation}
and $r_V$ is defined 
as in Eq.~(\ref{dRGTVain}) (Note that such cubic correction to the Newtonian potentials, $(r/r_V)^3$ 
has also been discussed in~\cite{Alberte:2010it} prior to the formulation of dRGT theory).

For the spherically symmetric static ansatz it is in fact possible to combine (\ref{EOMdRGTh0}) and (\ref{EOMdRGTp0}) to obtain a single 
quintic algebraic equation, say, on function $\tilde\mu$~\cite{Koyama:2011yg,Chkareuli:2011te},
\begin{equation}\label{dRGTmu}
\frac32\tilde\mu+\frac{3\tilde\alpha}{2\Lambda_3^3}\tilde\mu^2 +\left(\frac{\tilde\alpha^2}2+\frac{\tilde\beta}3\right)\frac{\tilde\mu^3}{2\Lambda_3^3}
-\frac{\tilde\beta^2\tilde\mu^5}{96\Lambda_3^3} = \frac{M_Pr_S}{r^3}\left(1 - \frac{\tilde\beta\mu^2}{4\Lambda_3^3}\right),
\end{equation}
and the potentials of the metric are expressed in terms of $\tilde\mu$, as
\begin{equation}
	\Psi' = \Phi' = \frac{M_P r_S}{r^2} - \frac{\tilde\beta}{24\Lambda_3^6}r\tilde\mu^3.
\end{equation}
Note again that Eq. (\ref{dRGTmu}) is similar to the equation found for general NLFP theory in decoupling limit, (\ref{FIbis}), (\ref{QDL}), with the same difference as above (and the same reason behind): it is algebraic and not differential.
It is worthwhile to mention that since the equation for $\tilde\mu$ is  in general of  order 5, there exist several branches of solutions, 
and depending on the parameters of the quintic equation, some of them yield real solutions. 
Moreover, as it was found in \cite{Koyama:2011yg}, there is a branch for which inside the Vainshtein radius 
the solution does not approach GR; in fact gravity becomes weaker when approaching the source. 
It turns out, however, that these solutions do not match a flat asymptotic behaviour~\cite{Sbisa:2012zk}. The study of the dRGT model in DL for all parameters of the theory and different asymptotic behavior has been presented in~\cite{Sbisa:2012zk}.

\subsection{The Vainshtein mechanism in complete Non Linear Fierz-Pauli theories:  recovering General Relativity away from the decoupling limit} 

The discussion of the previous section suggest that in untruncated massive gravity theories (and a class of scalar tensor theories with 
 non-linear kinetic self-interaction), the Vainshtein mechanism works. 
The results exposed so far for massive gravity, however, have been obtained in an approximation scheme:
the decoupling limit, in which most of the interactions were neglected
and it is possible to separate different helicities. 
One could hence ask how these results extend to the complete theory, i.e. when one does not truncate to the leading non linear interactions among these helicities.
Results, mainly based on numerical works, have been obtained addressing this question as we are going to explain here.

Static, spherically symmetric, and asymptotically flat numerical solutions of NLFP with a source were first studied in \cite{Damour:2002gp} which found only singular solutions and hence concluded that the Vainshtein mechanism was not correct. This was later re-examined for the same theories, using more sophisticated numerical methods, in~\cite{Babichev:2009jt,Babichev:2010jd}, where, for the first time, solutions featuring the Vainshtein recovery were found. 
The solutions  show a recovery of the Schwarzschild solution of GR via the Vainshtein mechanism 
for the potential of the form~(\ref{S3}).  The ansatz for the physical and the fiducial metrics reads respectively,
\begin{equation}
\eqalign{ 
g_{\mu \nu}dx^\mu dx^\nu &= -e^{\nu(r)} dt^2 + e^{\lambda(r)} dr^2 + r^2 d\Omega^2  \; ,\\
f_{\mu \nu}dx^\mu dx^\nu &= -dt^2 + \left(1-\frac{r \mu '(r)}{2}\right)^2 e^{-\mu(r)} dr^2 + e^{-\mu(r)}r^2 d\Omega^2\;, }
\label{lammunu}
\end{equation}
so that the physical metric has  a form which is convenient for comparison with the GR Schwarzschild solution, 
while the metric $f_{\mu\nu}$ has a non-canonical form, even though it parametrizes a Minkowski space-time. The function $\mu$ appearing above is related to the previously introduced $\tilde{\mu}$ of the DL as $\tilde{\mu} = m^2 M_P \mu$. This ansatz, where both metrics are diagonal, is not the most general one for a static and spherically symmetric case (see e.g. \cite{Deffayet:2011rh}), however it is the one used to describe the Vainshtein mechanism. For this ansatz, the modified Einstein equations and the Bianchi identity (\ref{BIAN}) 
form a set of three independent equations. 
Supplied with the conservation equation for matter source 
(for simplicity, a smoothly distributed source is taken to be described by a perfect fluid with energy density $\rho$ and pressure $P$), 
a set of four quasilinear ODEs for the four independent functions $\lambda(r)$, $\nu(r)$, $\mu(r)$, $P(r)$, was obtained, 
subject to boundary conditions at infinity and at the origin.
By solving numerically the field equations, using both shooting and relaxation methods as well as analytic insights,
everywhere non singular asymptotically flat solutions $\{\nu,\lambda, \mu, P\}$ were  found in Refs. \cite{Babichev:2009jt,Babichev:2010jd}. These solutions were found there to exist for a large range of parameters of the theory and of the source, but such that the source is not too compact (i.e. $r_{\odot} > 5 r_{S}$)\footnote[2]{The numerical analysis presented in \cite{Babichev:2009jt,Babichev:2010jd} breaks down for higher compactness and so far no other results have been obtained in these cases. This is further discussed below.}. They are non singular and asymptotically flat and have the right boundary conditions at the origin $r=0$. They feature a recovery of GR at distances $r \ll r_V$, where is was also checked numerically that the first correction to the GR behavior agree with the analytic expressions obtained  by expanding around  the Schwarzschild solution. 
 They are also well approximated by the solutions obtained in the decoupling limit in the expected range of distances $r_S \ll r \ll m^{-1}$, including in
the linear regime at $r_V \ll r \ll m^{-1}$ where deviations from GR  (due to the presence of the scalar polarization also responsible for the vDVZ discontinuity) are found. 
At distances above $m^{-1}$ the solutions feature the exponential Yukawa falloff, which again agrees with the the analytic results found by linearization.
A  crucial feature (which was missed in Ref.~\cite{Damour:2002gp}) of the asymptotic solutions at infinity is that they are not uniquely defined by standard perturbation theory  (i.e. expanding the solution in the Newton constant, as started in (\ref{GNexp})), 
but  have non-perturbative hairs allowing to match sources at smaller distances \cite{Babichev:2010jd}. 

Besides the potential (\ref{S3}) other potentials were also studied \cite{Babichev:2010jd}, 
in particular those which were shown in \cite{Babichev:2009us} to possess only a non-Vainshtein like scaling at $r<r_V$ in the decoupling limit. 
According to the numerical investigations of \cite{Babichev:2010jd}, these solutions found in the decoupling limit 
do not seem to continue into non singular solutions of the full field equations. 
A new limit was also introduced, the weak-field approximation, which captures all the salient feature of the solution, including the Yukawa decay and the Vainshtein crossover. Although the theories that has been studied in Refs. \cite{Babichev:2009jt,Babichev:2010jd} suffer from pathologies, in particular the Boulware-Deser ghost problem, these references were the first to exhibit an actual and complete proof that the Vainshtein mechanism can actually work in a massive gravity theory.
These results were later confirmed by an independent numerical
integration by Volkov~\cite{Volkov}, and analogous results were obtained in the dRGT model (discovered later) by this author~\cite{Volkov:2012wp} 
as well as by another group~\cite{Gruzinov:2011mm}, 
which used the shooting method to exhibit solutions featuring the Vainshtein behavior at small radii and the linearized Fierz–Pauli behavior at large radii.
Another numerical investigation of spherically symmetric solutions in dRGT massive gravity was performed in~\cite{Brihaye:2011aa},
where, however, only the case  with a very large graviton mass, $r_S m\sim 1$, has been studied. A numerical study of the cubic galileon model coupled to gravity~\cite{Kaloper:2011qc} has shown that the Vainshtein mechanism works in this model, 
also when taking into account full non-linear gravity.

In the bi-gravity extension of dRGT model \cite{Hassanbimet}, numerical investigation of static spherically symmetric solutions has been carried out using in particular a multiple shooting method \cite{Volkov:2012wp} (see also Volkov's contribution in this volume). Asymptotically flat solutions which exhibit the Vainshtein mechanism of recovery of General Relativity at finite distances, in the presence of a matter source have been found. 

Note however that none of the found numerical solutions, either in a generic NLFP, or in dRGT model or its bimetric extension, have sources with high compactness.  Indeed, when one increases the density of the object, the numerics becomes unstable and singularities are found to appear \cite{Babichev:2010jd,Volkov}.
It is still not clear if those singularities are physical or if they can be attributed to numerical artifacts. In the first case, it would indicate that the solutions cease to exist even before reaching a black hole size.  On the other hand, one can show that the standard Vainshtein mechanism does not work for black holes. Indeed, Ref.~\cite{Deffayet:2011rh} shows using geometric arguments, that a geometry such as the one given by (\ref{lammunu}) where the physical metric $g$ would contain a black hole horizon, must necessarily stop at or before the horizon. This results does not rely on the field equations and hence applies to a variety of cases including generic NLFP theory, dRGT theory and its bimetric extensions. This raises the interesting issue of the end point of gravitational collapse in these theories, given in particular that black holes solutions with metrics that are not both diagonal (and hence with no standard Vainshtein mechanism) are known to exist there.  

\subsection{Time dependent and other situations with Vainshtein screening}
First, following our discussion about static cases, it is worth  stressing that although the Vainshtein mechanism successfully operates in these cases, it might not be enough to pass standard gravity tests in the solar system. Indeed, in a generic shift-symmetric k-mouflage model, {the Vainshtein mechanism does not appear to screen sufficiently the induced time variation of the Newton constant, whenever the scalar field has some cosmological dynamics setting the boundary conditions at infinity (this result, first obtained in \cite{Babichev:2011iz}, has been confirmed by explicit study of the Horndeski theory~\cite{Kimura:2011dc})}.  Since the variation of the Newton constant is stirringly constrained by the Solar system experiments, this puts severe constraints on the direct matter-scalar coupling. Another related aspect is that non trivial Galileon profiles have been found to appear even around black holes due to a cosmological time evolution of the scalar \cite{Babichev:2010kj}.

Another important time-dependent situation is that relevant for gravitational waves.
Because massive gravity has at least five propagating degrees of freedom, 
the gravitational radiation from binary pulsars should occur at higher rate. 
One, however, could also expect that the Vainshtein screening dump at least the radiation from the scalar degree of freedom.
As an example of a system possessing the Vainshtein mechanism, gravitational radiation from binary pulsars  has been studied in a Galileon model in Refs. \cite{deRham:2012fw,deRham:2012fg}.
In case of the cubic galileon it was found that the radiation is indeed suppressed, although not as much as the static fifth force. This is because, firstly, monopole and dipole radiation were found to exist,  and secondly, because the suppression factor is larger than one could naively guess.
Indeed, one could have expected that because the static scalar field inside the Vainshtein radius is suppressed by 
a factor $(r/r_V)^{3/2}$\footnote{To get the static spherically symmetric solution in the case of the cubic galileon, one should take Eq.~(\ref{Emu}) without the cubic term. Then instead of $(r/r_V)$ and $(r/r_V)^2$ factors, in Eq.~(\ref{dRGTVAINSHTEIN}), one obtains a $(r/r_V)^{3/2}$ suppression of the scalar field.}, the radiation is suppressed by the same factor, in comparison to the  Brans-Dicke theory.  However, due to the presence of another scale in the problem, the orbital period $\Omega_P$,
the suppression turns out to be milder, namely by factor $\left(\Omega_P r_V\right)^{-3/2}$~\cite{deRham:2012fw}. In case of general galileon the study of radiation becomes more subtle, 
since the perturbation scheme breaks down for physically interesting situations \cite{deRham:2012fg}.  For the bi-gravity extension of dRGT model, a recent discussion  points out the possibility to detect a graviton mass via observations by gravitational wave detectors~\cite{DeFelice:2013nba}.
 A possibility to detect gravitational waves from massive gravity by means of the stochastic gravitational wave observations has also been studied in \cite{Gumrukcuoglu:2012wt}.

Other situations of great interest, where the Vainshtein mechanism has to be taken into account, are cosmology and many body problems (see in particular Refs \cite{Others}). However, we will not discuss this here and refer the reader to the contributions of Volkov and Koyama in this volume.

\section{Conclusions}
The Vainshtein mechanism has been the subject of many recent investigations and became a central piece in many theories of modified gravity. Although proposed more than 40 years ago, it is only very recently that the original idea of Vainshtein was shown explicitly to be correct. Several questions remain however open. Among the most pressing ones evoked above, we could mention the nature of the end point of gravitational collapse of a star or the way the Vainshtein mechanism can hide sufficiently time variations of the Newton's constant in the solar system. More generally, very little is known about time dependent cases. To conclude, we would like also to stress that the Vainshtein mechanism relies crucially, as it should appear clear from this paper, on being able to take into account non linear strong self-interactions, which are in general non renormalizable, in some sector of the considered theory. 
This calls for an understanding of the UV completion of these theories that is currently lacking.
Various possibilities have been put forward concerning this questions in the various theories concerned including the DGP model \cite{STRINGDGP,Adams:2006sv,CLASSICALIZATION}, but this definitely deserves more investigations if one wants to put these theories on a firm footing.

\ack 
The work of E.B. was supported in part by grant FQXi-MGA-1209 from the Foundational Questions Institute. C.D. thanks the IPMU Tokyo for its hospitality while part of this work has been completed.

\end{document}